\documentclass[journal]{IEEEtran}

\usepackage{graphicx}
\usepackage{subfigure}
\usepackage[justification=centering]{caption}
\usepackage{multirow}
\usepackage{booktabs}
\usepackage{algorithm}
\usepackage{hyperref}
\hypersetup{
    colorlinks,
    citecolor=black,
    filecolor=black,
    linkcolor=black,
    urlcolor=black
}

\usepackage[cmex10]{amsmath}

\usepackage[section]{placeins}
\usepackage{algpseudocode}
\usepackage{float}
\usepackage{translator}

\usepackage{array}

\usepackage{url}

\begin{document}
%
\title{Driving Simulator Platform for Development and Evaluation of Safety and Emergency Systems}
%
%

\author{Andr\'es E. G\'omez, Tiago C. dos Santos, Carlos M. Massera; \\Arthur de M. Neto$^{*}$, Denis F. Wolf
        
\thanks{Andr\'es G\'omez, Tiago dos Santos, Carlos Massera and Denis Wolf are with Sciences Institute of Mathematics and Computers, Mobile Robotics Laboratory, University of S\~ao Paulo, Av. Trabalhador S\~ao-Carlense, 400 - P.O. Box 668 - 13.560-970, S\~ao Carlos, Brazil
        {\tt\small $\{$phdaegh,tiagocs,massera,denis$\}$@icmc.usp.br}}%
\thanks{$^{*}$Arthur de M. Neto researcher is with the Terrestrial Mobility Laboratory, Federal University of Lavras, Engineering Department DEG / UFLA, mailbox 3037 CEP 37200-000, Lavras, Brazil
        {\tt\small arthur.miranda@deg.ufla.br }}%
}



\maketitle

\begin{abstract}

According to data from the United Nations, more than 3000 people have died each day in the world due to road traffic collision. Considering recent researches, the human error may be considered as the main responsible for these fatalities. Because of this, researchers seek alternatives to transfer the vehicle control from people to autonomous systems. However, providing this technological innovation for the people may demand complex challenges in the legal, economic and technological areas. Consequently, carmakers and researchers have divided the driving automation in safety and emergency systems that improve the driver perception on the road. This may reduce the human error. Therefore, the main contribution of this study is to propose a driving simulator platform to develop and evaluate safety and emergency systems, in the first design stage. This driving simulator platform has an advantage: a flexible software structure.This allows in the simulation one adaptation for development or evaluation of a system. The proposed driving simulator platform was tested in two applications: cooperative vehicle system development and the influence evaluation of a Driving Assistance System (\textit{DAS}) on a driver. In the cooperative vehicle system development, the results obtained show that the increment of the time delay in the communication among vehicles ($V2V$) is determinant for the system performance. On the other hand, in the influence evaluation of a \textit{DAS} in a driver, it was possible to conclude that the \textit{DAS'} model does not have the level of influence necessary in a driver to avoid an accident.  

\end{abstract}


\IEEEpeerreviewmaketitle

\section{Introduction}


During the last 20 years, in the area of Intelligent Transportation Systems (\textit{ITS}), the main objective has been to find new alternatives that help increment the levels of road safety. Some alternatives to solve this problem are:  \textit{i)} improve the road infrastructure for the drivers and \textit{ii)} develop safety/emergency systems for the vehicles. However, the high cost of the first alternative is a problem, mainly in developing countries where the rate of death by car crashes is high \cite{UN}. Therefore, the safety and emergency systems may be considered the best choice.

In this context, Sitavancová, in \cite{Sitavancová}, describes how the research in \textit{safety and emergency systems} covers the technology creation and new developments using techniques of several research fields (i.e., computer vision, control systems, and communications $V2X$ among others). These developments may enable faster responses to incidents; better guidance; collision avoidance; better energy use, and improve fleet management.

For more than three decades, the driving simulators have been an alternative to help in the development and evaluation of new safety and emergency systems. It is clear that a driving simulator cannot represent the real world with all circumstances. However, a simulator may be configured with multiple characteristics (i.e., modified car, visual system, motion system, sound, among others) that allow to best abstract the real world. 


The main contribution of this study is focused on the development of a driving simulator platform. This platform allows an open and flexible way; develop and evaluate prototypes of new safety and emergency systems; use different scenarios (i.e., where it is possible to interact with other vehicles) and the use of the $V2X$ adaptable communication networks.

Through this driving simulator platform, the users may have the possibility to adapt the structure with its tools (i.e., hardware and/or software) in the development or evaluation of new systems. This paper presents two examples of application of the driving simulator platform, which may contribute to improve the road safety.


The remainder of this paper is organized as follows. Section II reviews the state of art in driving simulation platforms; Section III explains the driving simulator platform architecture proposed. The experimental setup, results, and analysis from a development of a cooperative vehicle system, and the evaluation of $DAS'$ influence on a driver are presented in Sections IV and V. Finally, section VI provides conclusions and suggests future work.

\section{Related Work}

\subsection{History}

The idea of using simulators in driving training issues originates from the first flight simulator which was developed in the early 1910s \cite{Hassan}. This first simulator was created to be used for training pilots and for reducing operating costs required, compared to the use of real equipment \cite{Blana}. The widespread use of flight simulators inspired researchers to apply the same concept for road vehicles \cite{Hassan}. 


\subsection{Classification of driving simulators}

The significant difference among driving simulators is focused essentially in the characteristics of the components used in their fabrication. These components define the level of fidelity (i.e., \textit{fidelity} describes the degree to which the simulator's characteristics faithfully replicate the driving task \cite{Jamson}) in a driving simulator, and for this reason, there is a direct economic dependence. Therefore, the driving simulators can be classified as low, medium, and high-cost according to their acquisition cost. Sometimes they can also be referred to as low-level, mid-level, and high-level driving simulators \cite{Lang}.

\begin{figure*}[!htb]
\centering
\framebox{\includegraphics[width=9cm,height=6cm]{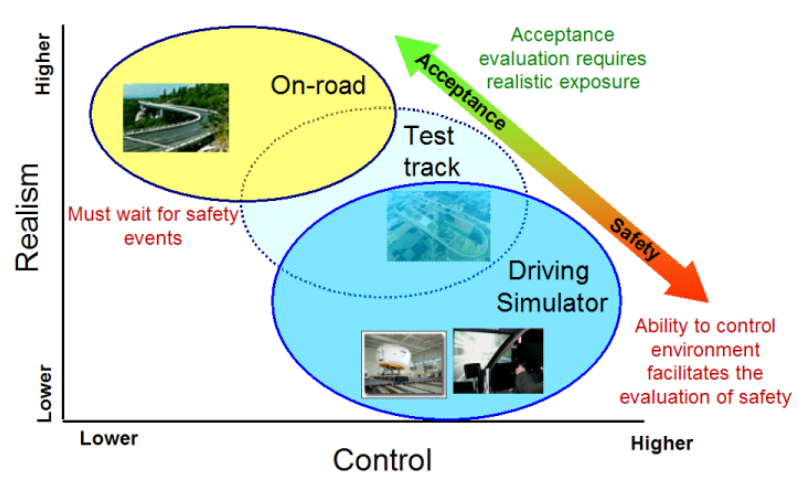}} 
\caption{Selection of the best \textit{test track} for an application, in function of the realism and the control. Taken from \cite{Smith} }\label{fig:testTrack}
\end{figure*}

It is important to consider that the relation of cost-benefit, in the driving simulators depends on the goal and on the level of details needed to represent the real world. In Figure \ref{fig:testTrack} Smith et al. emphasized that the best \textit{test track} in a simulation has to combine the realism with the controlled situations, in function of the acceptance and safety \cite{Smith}. For this reason, it is common to find driving simulators used in researches with a representative cost, regarding the use of a simulator for training.   

\subsection{Driving simulators used by the carmakers}\label{sec:car makers}

The car makers use driving simulators to test a new system on a vehicle and to analyze how the system could influence the driver. For this reason, the driving simulation platforms used by the carmakers have the highest possible fidelity. Some examples of driving simulators used by the carmakers are the ULTIMATE by $Renault’s$, Dynamic Driving Simulator by $BWM$, the Virttex by $Ford$, the Daimler AG driving simulator by $Mercedes$, and the driving simulator used by $Nissan$. 


\subsection{Driving simulators used in Research}\label{sec:research}

In researches, the driving simulators have focused on the analysis of drivers' behavior regarding some hazard situations, considering specific factors that could affect the driver (e.g. mobile phones, psychoactive substance use, test traffic signals, test driver assistance systems, among others). Considering \cite{Castro}, some variables measured by the driving simulators in researches are: \textit{i)} vehicle speed; \textit{ii)} vehicle place on road markings; \textit{iii)} distance from the vehicle in front; \textit{iv)} angle of the steering wheel; and \textit{v)} amount of pressure applied to the brake pedal. Therefore, the driving simulators used in researches have a high fidelity. In researches the nature of the experiments is always dynamics (i.e., it may be necessary to change some characteristics for the driving simulator platform), and for this reason, the cost and performance of the awaited results could be compromised. Therefore, when a driving simulator is used in researches, it is necessary that the system structure be re-configurable\cite{Hassan}. 

Nowadays, there are multiple driving simulator platforms to develop and evaluate safety and emergency systems. Some of them are commercial and are more prominent by its technical characteristics, such as: the exact dynamic model used in the vehicles; integration with other software tools; and the fidelity of obtained results. Moreover, some of these platforms are used by some car makers and research centers \footnote{\url{https://www.carsim.com/products/ds/index.php}}\footnote{\url{https://www.tassinternational.com/testimonials}}. However, these driving simulator platforms have considerable acquisition and maintenance costs. Furthermore, the commercial driving simulator platforms also have a closed code, which allows only a minimal tool adaptation. Among some commercial driving simulator platforms most known, it is possible to find: \textit{CarSim and TruckSim}, from \textit{Mechanical Simulation}; \textit{PreScan}, from \textit{TASS International}; and \textit{CarMaker}, from \textit{IPG Automotive GmbH}.  

There are also state of the art open-sourced driving simulator platforms, which, in addition to the economic benefit, allow the simulator platform to adjust to the needs of the user. However, one of the biggest drawbacks of these simulator platforms is that they do not provide users with adequate technical support. For this reason, they require a longer learning time. Among some open-source driving simulator platforms, there are: the \textit{TORSC} \cite{Wymann}, which is a highly portable multi-platform car racing simulation and used as an ordinary car racing game, as an $AI$ racing game and as a research platform; and the \textit{Delta3D} \footnote{\url{http://delta3dengine.org/}}, which is a game and simulation engine appropriate for a wide variety of simulation and entertainment applications.

The driving simulator platform presented in this study is an open-sourced tool, with a middle level of fidelity (See Section \ref{s:III}). Its objective is to contribute to the development of new safety and emergency systems, in the first stage of modeling. Similar to other open-sourced driving simulator platforms, this tool allows the users to configure its architecture by using different software tools (i.e., also open-source) that integrate it. The integration of the driving simulator platform with different software tools could be considered its main advantage. For example, this characteristic allows the integration of a vehicular communication model $V2V$ in the platform.     

Its main disadvantage is the fact that it does not count on an own dynamic model for the vehicles used in the simulations. In fact, the development of these dynamic models is one of our activities. In Section \ref{s:III}, the driving simulator platform architecture is presented. In this Section, it will also be described the current dynamic model used in vehicles' platforms.
%

\subsection{Advantages and disadvantages of driving simulators}

A major disadvantage of a driving simulator is that it is not able to represent the complexity of real road situations. For this reason, the validity and reliability of these tools are often questioned in the research area. However, despite of this disadvantage, research centers \cite{Rudin}-\cite{Yoshimoto} and car makers \cite{Aeberhard}\cite{Zeeb}\cite{Challen} have a driving simulator available based on its significant advantages, such as, facilitate the design, development, and test of their proposals. 


\subsection{Validation of driving simulators}\label{ss:val_DS}

The models used in any driving simulator have to pass through tests and validations regarding the system, entity or idea that is being represented. The objective is to obtain a level of confidence in the driving simulator \cite{Espie}. Engen in \cite{Engen} considers that the driving simulators have to guarantee that the results obtained can be replicated (i.e., at least in a high percentage) in a real field-test. However, it is not always possible to compare the results of a driving simulator with the results of a real field-test (i.e., generally, for economic and logistic limitations), therefore it is necessary to find other alternatives. Engen proposes to consult a larger array of sources in order to find such alternatives and evidence of validity to be included in research models. Evidence of validity can be confirmed according to behaviors or specific events obtained from the tests results of the driving simulators. 

\section{Driving simulator platform architecture}\label{s:III}

The previous section had as objective to consider some relevant aspects of the driving simulators and their applications. For example, the relation between the realism and the control of the simulated situations in this driving simulation platform could be located at the \textit{test track’s} circle's center point, as shown in Figure \ref{fig:testTrack}. The main reason for why this particular point was selected in the \textit{test track} is due to the lack of a higher level of realism in the models used in the current platform version (i.e., exact vehicle dynamics and environmental conditions of the simulation scenario, among others). Taking into account this lack of realism in the models of the current version, it is recommended the use of this simulation platform only in the first stage of the system development. The obtained result using the presented driving simulator platform could help the platform users to consider a test of one's system in real field-tests, as it was made in the Subsection \ref{ss:Validation}. 
\begin{figure}[thpb]
\centering
\framebox{\includegraphics[width=8cm,height=5cm]{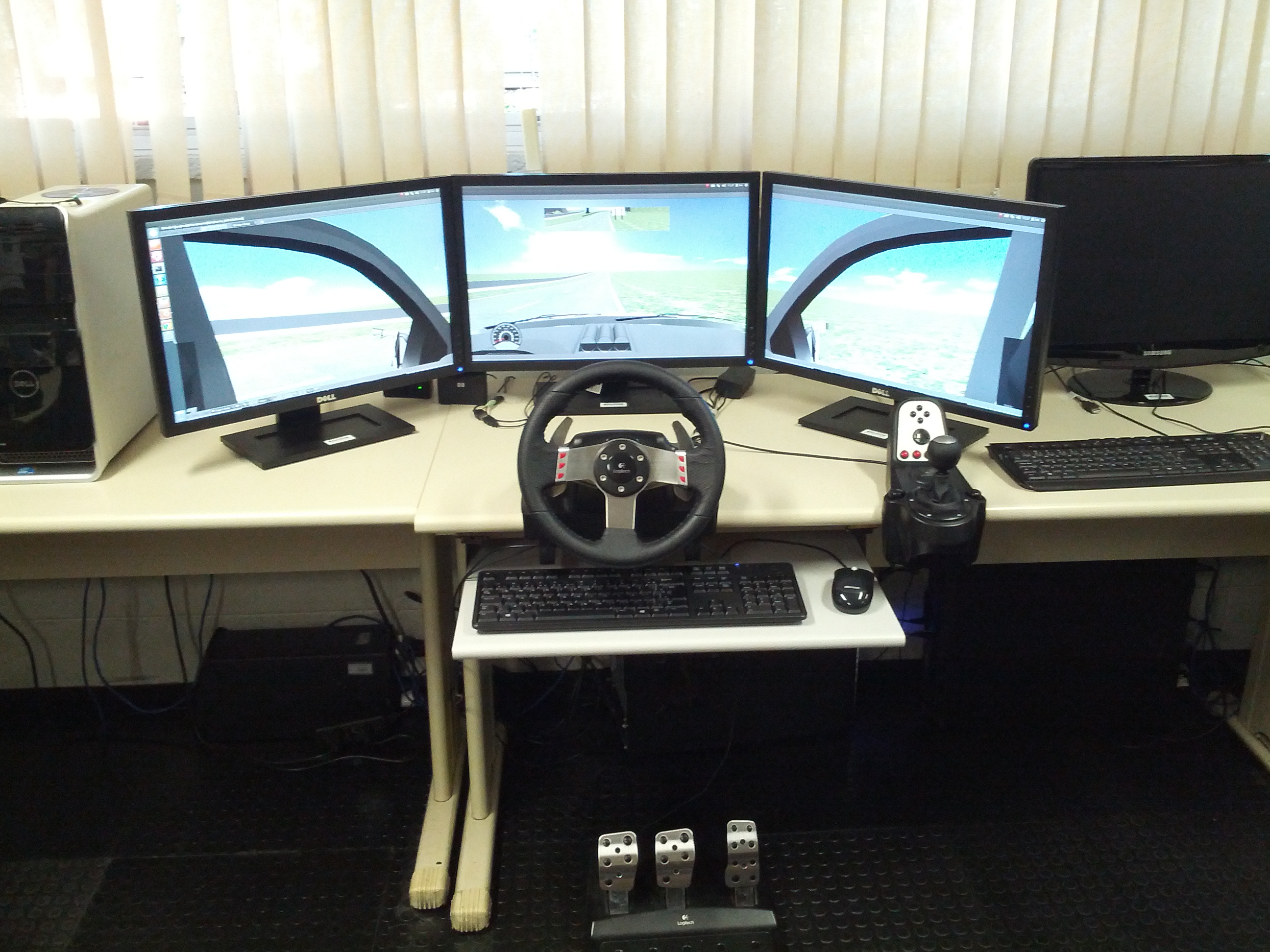}}
\caption{Hardware used in the driving simulator platform}
\label{fig:platform}
\end{figure}

The driving simulator platform proposed, is divided in three main parts: \textit{i)} a generic simulator for academic robotics, \textit{ii)} a middleware for robotic applications, and \textit{iii)} a communications simulator. The generic simulator for academic robotics used is the Modular Open Robots Simulation Engine $MORSE$ (i.e., the version used in this study was morse 1.2.2). $MORSE$ is the main tool for the driving simulator platform. As described by Echeverria et al. in \cite{Echeverria2011}, $MORSE$ is a free and open-sourced software tool, which can be adapted to the requirements of a particular system. $MORSE$ has a large number of sensors, which can be used in the development and evaluation of the safety and emergency systems. Furthermore, $MORSE$ proposes not only some methods to alter the input and output data (i.e., through modifiers) but also allows the creation of new sensors\footnote{\url{https://www.openrobots.org/morse/doc/stable/contributing.html}}. For example, in this proposal it was created the Network Interface Communication $NIC$ sensor to emulate a communication vehicular card.        

\begin{figure}[thpb]
\centering
\framebox{\includegraphics[height=240px]{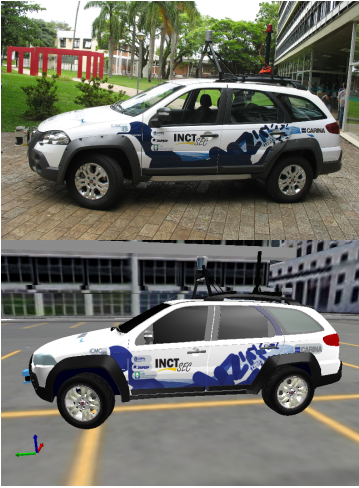}}
\caption{The 3D simulation model of CaRINA 2 \cite{carina}.}
\label{fig:veiculos}
\end{figure}

The simulations in $MORSE$ are based in $Blender$, which is a tool with the main purpose of creating images and $3D$ computer animations through a high level of graphic detailing (i.e., the version used was blender-2.73a). This characteristic of $MORSE$ allows the creation of different $3D$ vehicle models (see Figure \ref{fig:veiculos}) with some dynamic characteristics (see Subsection \ref{ss:dynamic}). With $Blender$ it is also possible to create virtual scenarios for the simulations. In this proposal, were considered realistic scenarios to create virtual scenarios, using GPS coordinates, which were plotted in the $Blender$ version used.   

In the driving simulator platform, it is also used the Robot Operating System $ROS$ \cite{ROS2009}, a middleware used in robotic, which has, as the main feature, libraries and tools to help software developers create robot applications (i.e., the version used was indigo). Given that the main tool of the driving simulator platform, $MORSE$, lets an easy integration with $ROS$, it is possible to get data of each one of the sensors in every vehicle during a simulation. Through $ROS$, it is possible to implement models (i.e., see Subsections \ref{cap:control} and \ref{ss:mDAS}).

Finally, the communications simulator used in the proposed driving simulator platform architecture is $NS-3$ (i.e., the version used was ns-3.19). With $NS-3$ it is possible to develop codes using the programming languages \textit{C++} or \textit{python}. Besides being open-source, $NS-3$ also contains some tools for analysis of networks (e.g., $pcap$). In $NS-3$ is possible to also find different models for computer networks (e.g., propagation and wifi among others). For example, in this study it was used the model Wireless Access in Vehicular Environments $WAVE$, to simulate the vehicle to vehicle communication $V2V$, used in sections \ref{a1} and \ref{a2}. In the simulation using $NS-3$, the vehicular communication $V2V$ has a node (i.e., in $NS-3$ a node is a computer device that connects with a network) which represents a vehicle of the simulation executed in $MORSE$. Figures \ref{fig:arq1} and \ref{fig:arq2} present two different examples of the architecture's configuration proposed in this section.      
In Figures \ref{fig:arq1} e \ref{fig:arq2} is possible to observe that the three main parts of the driving simulator platform architecture communicate with each other by using two $sockets$ (i.e., A socket\footnote{The ${Java}^{TM}$ Tutorials} is a software endpoint that establishes bidirectional communication between a server program and one or more client programs). In the proposed driving simulator platform architecture, $NS-3$ requests information to $MORSE$ through the $Socket-SS$. When $MORSE$ receives the request, $MORSE$ responds to $NS-3$ with a determined kinematic package data of the vehicle that has requested it. The answer of $MORSE$ is sent using the $Socket-SS$. Then, $NS-3$ receives the kinematic package data sent by $MORSE$, and directs the information to the correspondent node. Finally, the information received in $NS-3$ is sent to $ROS$ using the $Socket-MS$, with the objective to be used in a model. Data received by $NS-3$ are considered package in network simulation. This process is used by each node during all simulations. 

\subsection{Vehicular dynamic model}\label{ss:dynamic}

The actual Vehicular Dynamic Model (\textit{VDM}) used in this driving simulation platform is based on the $Bullet$ $physics$ engine adopted for $Blender$. $Bullet$ $Physics$ is an open-sourced library, written in portable \textit{C++}, used for professional collision detection and for rigid and soft body dynamics. This library is primarily designed for the use in games, visual effects, and robotic simulation \cite{Coumans}. 

$Bullet’s$ approach to a vehicle controller is called a \textit{Raycast Vehicle}. A \textit{raycast} vehicle works by casting a ray for each wheel. By using the ray’s intersection point, we can calculate the suspension length and, consequently, the suspension force that is then applied to the chassis, keeping it from hitting the ground. The friction force is calculated for each wheel where the ray contacts the ground. This is applied as a sideways and forwards force \cite{Blender}. A detailed example of the entire implementation of the \textit{VDM} used in the vehicles of this driving simulator platform can be found at \url{http://www.tutorialsforblender3d.com/Game_Engine/Vehicle/Vehicle_2.html}.  

\subsection{Validation of the driving simulator platform}\label{ss:Validation}

In this proposal, the driving simulator platform validation is based on a study, presented by Filho et al in \cite{Massera}, about the simplification of the amount of control system parameters for the navigation of an Autonomous Vehicle ($AV$). Regarding validation, in this study, the assessment of the lateral control law in the driving simulator platform (i.e., it was used the CaRINA 2 $Blender$ model, as seen in figure \ref{fig:veiculos}) is compared to the assessment of the same lateral control law used in the real testbed CaRINA 2 \cite{carina} in two field-tests. 

The driving simulator platform was used during the design and prototyping of the lateral control law. In the CaRINA 2 $3D$ model, were not considered the longitudinal forces applied in the vehicle. In order to obtain results as close to reality as possible, Filho et al considered some errors (e.g., steering encoder errors, IMU orientation errors, localization errors and steering speed limitation) in the CaRINA 2 $Blender$ model. The objective was to experiment with different controller setups. 

\begin{figure}[th]
\centering
\subfigure[DSP]{\framebox{{\includegraphics[width=8cm,height=5cm]{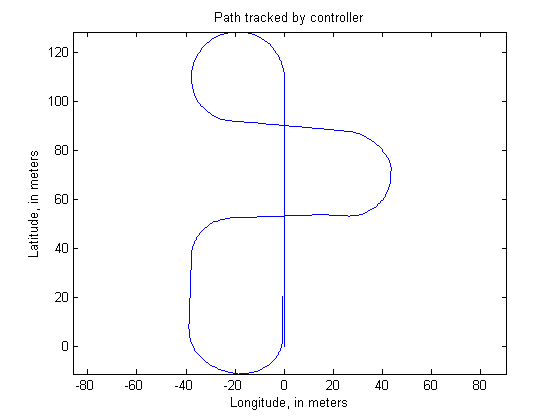}}}}
\subfigure[FOT1]{\framebox{{\includegraphics[width=8cm,height=5cm]{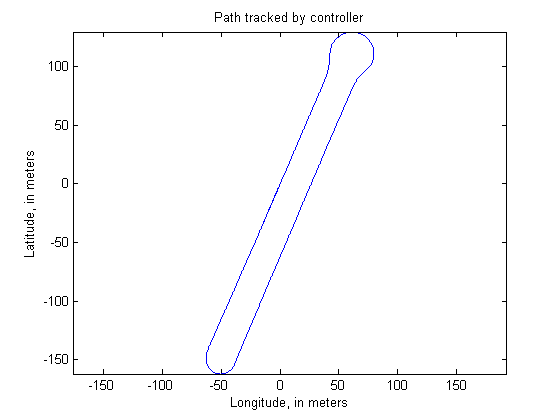}}}}
\subfigure[FOT2]{\framebox{{\includegraphics[width=8cm,height=5cm]{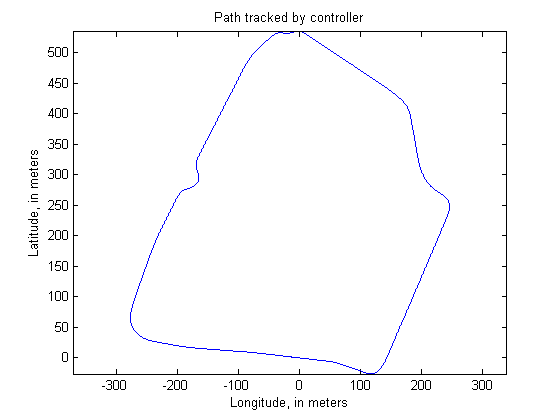}}}}
\caption{Tracks used in the Driving Simulator Platform (\textit{DSP}), Field of Test 1 (\textit{FOT1}) and Field of Test 2 (\textit{FOT2}). Taken from \cite{Massera} }\label{fig:validation}
\end{figure}

Regarding the lateral control law assessment in the driving simulator platform, Filho et al considered a complex track in order to acquire the best result. For this reason, the simulated track, showed in Figure \ref{fig:validation}a, is $400$ $meters$ long, with closed and opened turns, which allowed the reproduction of different  behaviors of the vehicle. Although this track does not exist in reality it has some characteristics of the real fields selected for the tests (e.g., Field of Test 1 (\textit{FOT1}) and Field of Test 2 (\textit{FOT2})). 

The \textit{FOT1} is a long loop track of $600$ $meters$ in length. In this trajectory, there are one-way asphalt roads with two lanes at all times, an accentuated left turn and a roundabout (see Figure \ref{fig:validation}b). Additionally, the \textit{FOT2} is a test track with a loop $1.6$ $kilometers$ long, going through $4$ roundabouts, with asphalt roads of one single two-way lane and two one-way lanes (see Figure \ref{fig:validation}c). 
   
Regarding the lateral control evaluation, the autonomous vehicle traveled around (i.e., in the driving simulator platform, as on both \textit{FOT}) the track selected for each case (i.e., \textit{DSP}, \textit{FOT1} and \textit{FOT2}). For several times, it was considered an average speed to get the cross track error and the orientation error in the control proposed. 

Table \ref{tv} shows the average results of the tests done on different tracks (i.e., \textit{DSP}, \textit{FOT1} and \textit{FOT2}) with their values of cross track error and orientation error (i.e., \textit{C. error} and \textit{O. error}). It is possible to see that the cross track error in every test is practically the same. However, in the orientation error, the difference among the driving simulation platform and the other two field-tests is considerable. The main reason of this error is presented in the beginning of this section. Therefore, it is considered that this driving simulator platform has a medium fidelity.     

\begin{table}[t]
\caption{Results of the lateral control evaluation}
\label{tv}
\begin{tabular}{|c|c|c|c|c|c|c|c|}
\hline
\multirow{2}{*}{\textbf{Tracks}} & \multirow{2}{*}{\textbf{Laps}} & \multicolumn{2}{c|}{\textbf{Speed (Km/h)}} & \multicolumn{2}{c|}{\textbf{C. error (m)}} & \multicolumn{2}{c|}{\textbf{O. error ($^{o}$)}} \\ \cline{3-8} 
                                 &                                & \textbf{$\mu$}     & \textbf{Max}     & \textbf{$\mu$}             & \textbf{$\sigma$}            & \textbf{$\mu$}             & \textbf{$\sigma$}             \\ \hline
\textbf{DSP}                     & 5                              & 31.3                 & 90                  & 0                         & 0.0808                 & 0.6327                    & 2.0173                  \\ \hline
\textbf{FOT1}                    & 3                              & 20.4                 & 28.8                & 0                         & 0.0971                 & -1.0397                   & 1.0397                  \\ \hline
\textbf{FOT2}                    & 3                              & 23.9                 & --                  & 0                         & 0.088                  & 0.9225                    & 1.0544                  \\ \hline
\end{tabular}
\end{table}

\section{Cooperative vehicle systems}\label{a1}

The variety of applications of the cooperative vehicle systems and their developments are also increasing rapidly \cite{Passchier2013}. For instance, the Cooperative Adaptive Cruise Control system ($CACC$) is an enhancement of the Adaptive Cruise Control system ($ACC$) that uses inter-vehicle communication to perform safe cruising at shorter inter-vehicle distance \cite{Omae2014}. In order to validate the proposed driving simulator platform architecture, a $CACC$ was used since it requires both $V2V$ communication and the simulation of multiple agents. A state-of-the-art controller based on \cite{Arem2006} was used.

\subsection{Control}\label{cap:control}

Given that $ [\Delta x_{i}, \Delta v_{i}, \Delta a_{i}]^{T} $ are the distance, relative speed and relative acceleration between the $i$-th and the $(i-1)$-th vehicle of a platoon, respectively, $ \Delta x_{i}^{safe} $ is the safety distance between vehicles given by

\begin{equation}
\Delta x_{i}^{safe} = h v_{i} + d
\end{equation}

where $ h $ is the headway time and $ d $ is the desired distance for a stopped vehicle. From the desired safe distance, it is possible to define the control error ${e}_{i}$:

\begin{equation}
\mathbf{e}_{i} = \Delta x_{i} - \Delta x_{i}^{safe}
\end{equation}

and its derivative ${\dot{e}}_{i}$:

\begin{equation}
\mathbf{\dot{e}}_{i} = v_{i-1} - v_{i} - h a_{i}
\end{equation}

Assuming that the acceleration is directly controllable and the jerk is unbounded, a $CACC$ can be defined as a PD controller

\begin{equation}
\Delta a_{i} =K_{p} (\Delta x_{i} - h v_{i} - d) + K_{d} (\Delta v_{i} - h a_{i})
\label{control_law}
\end{equation}

However, implementing it as a pure PD controller term would cause instability depending on the operation frequency since the current acceleration would be based on the previous one. In order to avoid this scenario Equation \ref{control_law} can be rewritten as:

\begin{equation}
a_{i} = \frac{a_{i-1} + K_{p} (\Delta x_{i} - h v_{i} - d) + K_{d} (\Delta v_{i})}{1 + K_{d} h}
\label{control_law_2}
\end{equation}

\subsection{Driving Simulator Platform Configuration}\label{cap:DSPC1}

\begin{figure}[thpb]
\centering
\framebox{\includegraphics[width=8cm,height=5cm]{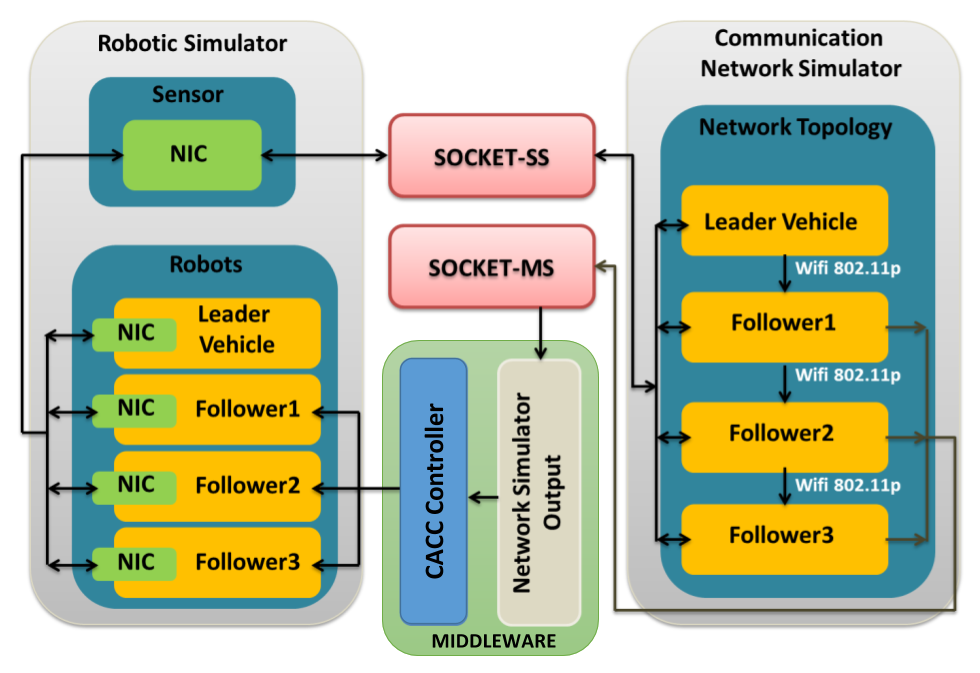}}
\caption{Driving simulator platform configuration used in the cooperative vehicle system.}
\label{fig:arq1}
\end{figure}%
Figure \ref{fig:arq1} presents the configuration of the driving simulator platform, used for the implementation and evaluation of the control’s model, present in the Subsection \ref{cap:control}. In the configuration of its architecture (i.e., which was first presented in \cite{Gomez2014}) it is possible to observe the three main parts of the driving simulator platform, described in Section \ref{s:III}. In Figure \ref{fig:arq1}, the rectangle \textit{Robotic Simulator} represents the $MORSE$ simulator. In that rectangle is possible to observe the vehicles (i.e., robots Leader, Follower1, Follower2, Follower3) used in the validation of the $CACC$ control (see Section \ref{ss:CACC_Validation}). In each vehicle simulated in $MORSE$, sensors (i.e., $GPS$, $velocity$, $pose$) were added to obtain data about velocity, position, and acceleration of the vehicles. Each information were used in the control model.        

In Figure \ref{fig:arq1} is also possible to observe that each vehicle in \textit{Robotic Simulator} uses a sensor Network Interface Communication \textit{NIC}. The sensor  \textit{NIC} creates a message with different data (i.e., ID vehicle, position (x,y,z), yaw, velocity, and acceleration) of each simulated vehicle in $MORSE$.  This message is shared with the rectangle \textit{Communication Network simulator} that represents the $NS-3$ simulator in Figure \ref{fig:arq1}. The sending of messages between the \textit{Robotic Simulator} and \textit{Communication Network simulator} rectangles is done through the use of the \textit{SOCKET-SS}. The sockets' function shown in Figure \ref{fig:arq1} is detailed in the last paragraph of the driving simulator platform architecture description, before Subsection \ref{ss:dynamic}.       

The communications model of vehicle to vehicle $V2V$, implemented in the communications simulator $NS-3$, consists in a precedent communication’s topology proposed by Y. Chen in \cite{Chen2013}. The $V2V$ communication is developed using four nodes, which are represented by each one of the vehicles used in $MORSE$ (i.e., Leader, Follower1, Follower2, Follower3). Each node in $NS-3$ has a physical layer $802.11p$ Wi-Fi, a $MAC$ layer and a definition of the hardware’s physical properties. Moreover, for each node in $NS-3$ was assigned an $IP$ address, with the purpose of sending data packages between nodes. When the communications simulator $NS-3$ obtains a message shared by the sensor \textit{NIC} in $MORSE$, the $V2V$ communications model, sends the obtained message to the corresponding vehicular node. This process is repeated throughout the system validation $CACC$.     

After a node receives a message, through the communications model in $NS-3$, it resends the message to the rectangle \textit{MIDDLEWARE}, which represents $ROS$, using the \textit{SOCKET-MS}. In this moment, it begins the execution of the control model presented in Subsection \ref{cap:control}. As it is described in the Section  \ref{s:III}, $ROS$ is a middleware that allows the implementation of models. It is divided into a set of nodes, and inside of this set there is one main node called $Master$. The $Master$ node initializes $ROS$ and allows the others nodes (i.e., these nodes are pieces of software) to communicate with each other. The $ROS'$ nodes are different from the nodes used in $NS-3$.       

In the validation of the control’s model with the driving simulator platform, seven nodes were implemented in $ROS$:  \textit{i)} one node to get the sent messages through the \textit{SOCKET-MS} (i.e., represented by \textit{Network simulator Output} inside of rectangle \textit{Middleware});  \textit{ii)} and a pair of nodes  \textit{CACC controller} (see Figure \ref{fig:arq1}) for each of the three $Follower$ vehicles, with the purpose of determining the behavior of each $Follower$ vehicle, based on the obtained messages from the node \textit{Network simulator Output}. Finally, the control's model effect can be observed in each $Follower$ vehicle throughout the execution of the simulation in $MORSE$. 

\subsection{Results}\label{ss:result}

As a validation for the driving simulator platform proposed in this study, two experiments were performed with the $CACC$ control described in Subsection~\ref{cap:control}. In every experiment we assumed that the vehicles had access to their own position and velocity information. A video with the execution of a test is available in the website \url{https://www.youtube.com/watch?v=WQsRR9ajiPc}.

\subsubsection{Delay effect on the CACC system}\label{subcap:exp1}

In the first experiment, two vehicles were used. The follower vehicle must keep a safe distance from the leader vehicle using the longitudinal control law described in Section~\ref{cap:control}. The purpose of this experiment was to evaluate the effect of the communication delays on the control of the follower vehicle. The scenario used for the experiments was an open space with no obstacles. 

\begin{figure}[thpb!]
\centering
\subfigure[Velocity in the $CACC$]{\framebox{{\includegraphics[width=7.5cm,height=4.3cm]{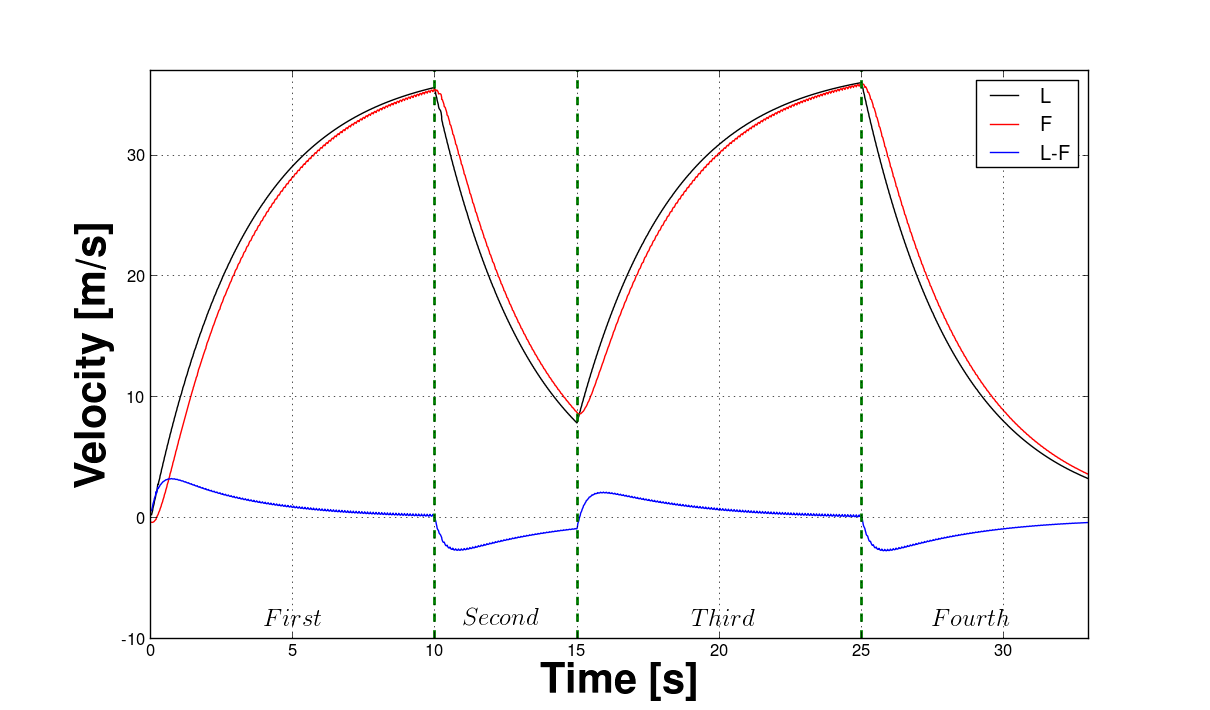}}}}
\subfigure[ the histograms of the velocity difference (L-F)]{\framebox{{\includegraphics[width=7.5cm,height=4.3cm]{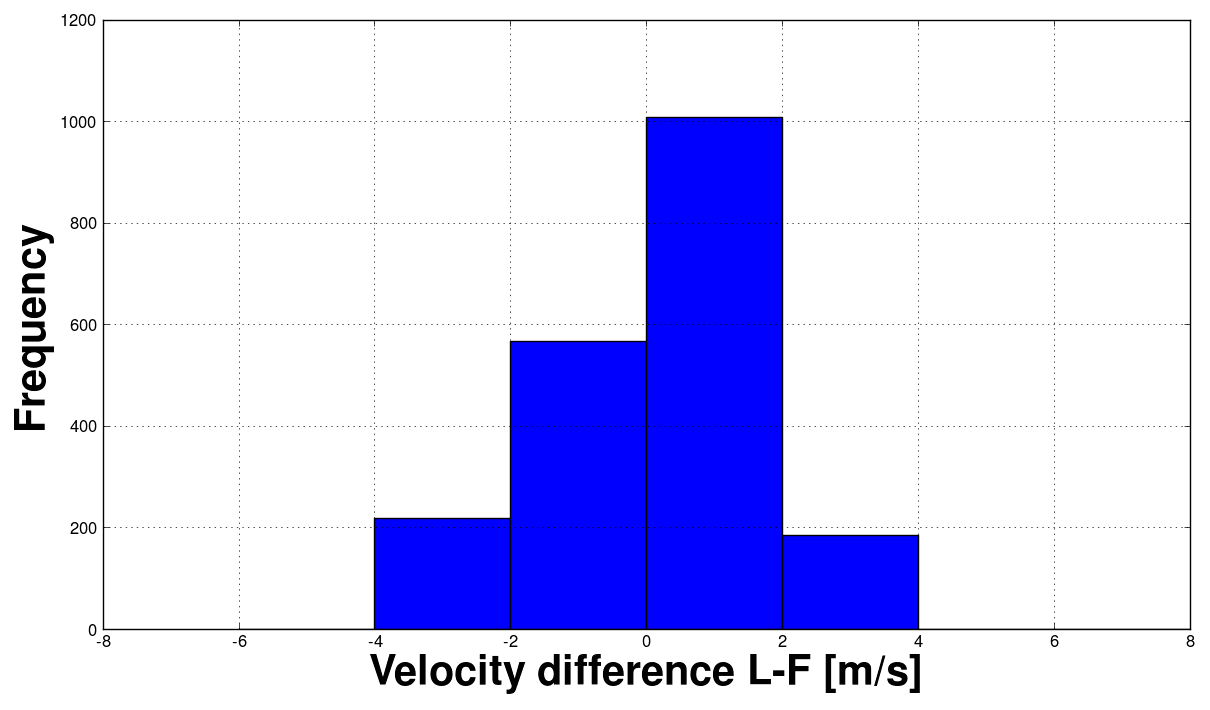}}}}
\caption{Leader vehicle Velocity (\textit{L, black}), Follower vehicle velocity (\textit{F, red}), velocities variation (\textit{L-F, blue}), communication delay 0.01 s.}
\label{fig:delay01}
\end{figure}
\begin{figure}[thpb!]
\centering
\subfigure[Velocity in the $CACC$]{\framebox{{\includegraphics[width=7.5cm,height=4.3cm]{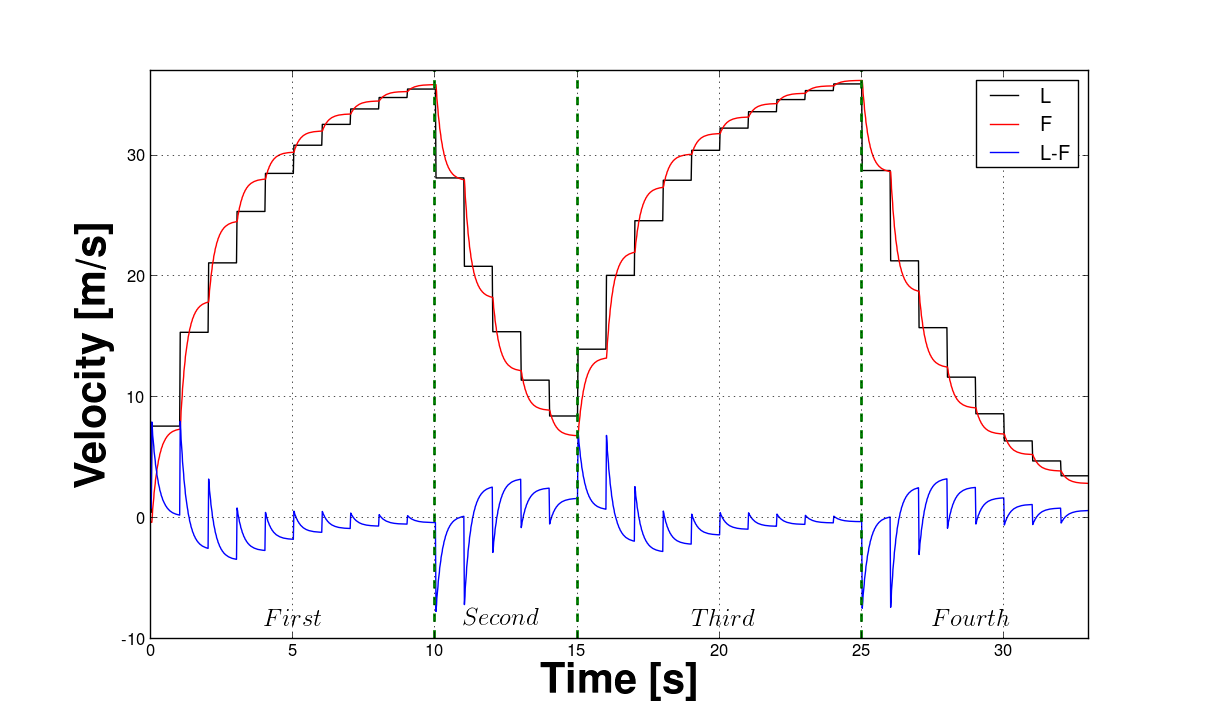}}}}
\subfigure[ the histograms of the velocity difference (L-F)]{\framebox{{\includegraphics[width=7.5cm,height=4.3cm]{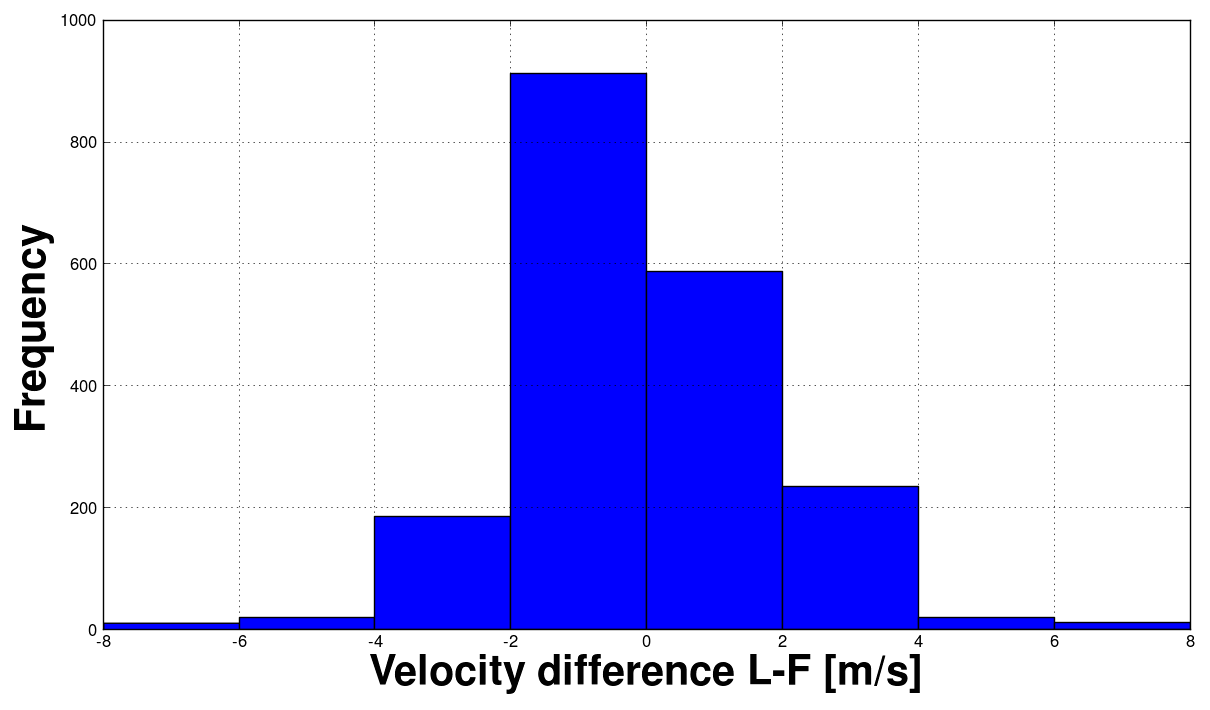}}}}
\caption{Leader vehicle Velocity (\textit{L, black}), Follower vehicle velocity (\textit{F, red}), velocities variation (\textit{L-F, blue}), communication delay 1.0 s.}
\label{fig:delay10}
\end{figure}

%
The tests have been performed using different wireless communication delays of $0.01$ \textit{s}, $0.1$ \textit{s}, $0.5$ \textit{s}, and $1.0$ \textit{s}. Figures \ref{fig:delay01} and \ref{fig:delay10} show the velocity of both vehicles obtained from the network simulator and the histogram of the difference in the velocity of the two vehicles, directly from the output of the $NS-3$. The total time of the experiment was $33$ \textit{s}. From $0$ \textit{s} to $10$ \textit{s}, the leader vehicle accelerated from $0$ to $37$ \textit{m/s}; from $10$ \textit{s} to $15$ \textit{s}, the velocity of the leader decreased to $8$ \textit{m/s}; from $15$ \textit{s} to $25$ \textit{s}, it increased again to $37$ \textit{m/s}; and from $25$ to $33$ \textit{s}, the leader decelerated until it achieved $3$ \textit{m/s}. Figures \ref{fig:delay01}a and \ref{fig:delay01}b show the velocity profile and histogram, respectively, of vehicle's velocity obtained from the network simulator with a $0.01$ \textit{s} communication delay. We can notice that the follower's velocity follows very close the leader's velocity. Meanwhile, Figure \ref{fig:delay10}a and Figure \ref{fig:delay10}b present the velocity profile and histogram, respectively, from the simulation with a $1.0$ \textit{s} communication delay. The higher communication delay results on a non-continuous velocity profile. Also the histogram presented in Figure \ref{fig:delay10}b shows a larger variance in the velocity difference when compared to the data presented in Figure \ref{fig:delay01}b.

Table \ref{t2} shows the average results of the velocity difference of both vehicles obtained from the network simulator with communication delay of $0.01$ \textit{s}, $0.1$ \textit{s}, $0.5$ \textit{s}, and $1.0$ \textit{s}. It is possible to notice that the maximum velocity variation increases with higher delay values.  

\begin{table}[t]
\caption{Statistics features of L-F velocity variations due to delay times in communication.}
\label{t2}
\begin{center}
\begin{tabular}{|c|c|c|c|c|}
\hline
\textbf{Delay (s)} & \textbf{$\mu$ (m/s)} & \textbf{$\sigma$ (m/s)} & \textbf{$\sigma^{2}$ $(m/s)^{2}$} & \textbf{Max. variation (m/s)} \\ \hline
0.01               & 0.0630             & 1.4620                & 2.1374                                  & 3.276                    \\ \hline
0.1                & 0.0540             & 1.2684                & 1.6089                                  & 3.582                    \\ \hline
0.5                & 0.0263             & 1.1926                & 1.4224                                  & 6.091                    \\ \hline
1.0                & -0.0121            & 1.7828                & 3.1786                                  & 8.007                    \\ \hline
\end{tabular}
\end{center}
\end{table}%

\subsubsection{CACC system validation}\label{ss:CACC_Validation}

In the second experiment four vehicles were used (one leader $L$ and three followers $F1$, $F2$, and $F3$) to evaluate the $CACC$ performance. Each follower vehicle only communicated to the vehicle immediately ahead of it.

\begin{figure}[thpb]
\centering
\subfigure[Velocity in the $CACC$ system]{\framebox{{\includegraphics[width=7.3cm,height=4cm]{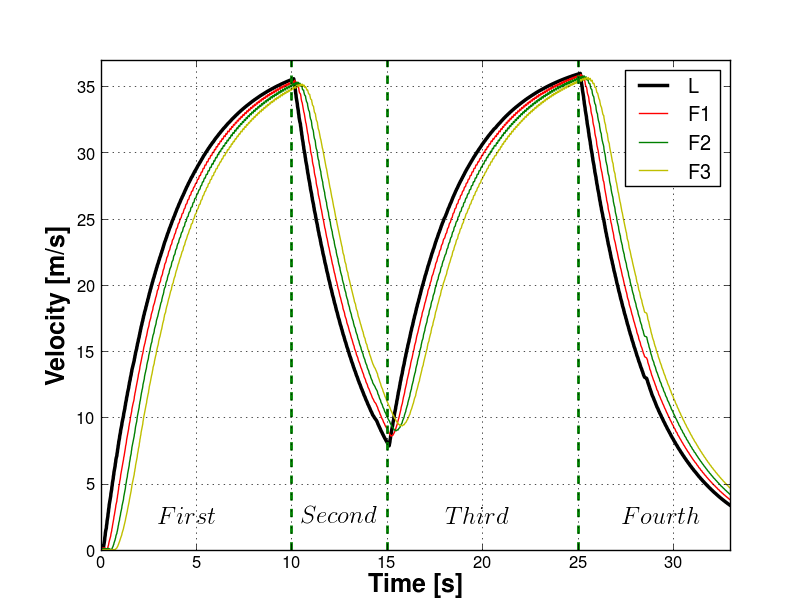}}}}
\subfigure[ Distance among vehicles]{\framebox{{\includegraphics[width=7.3cm,height=4cm]{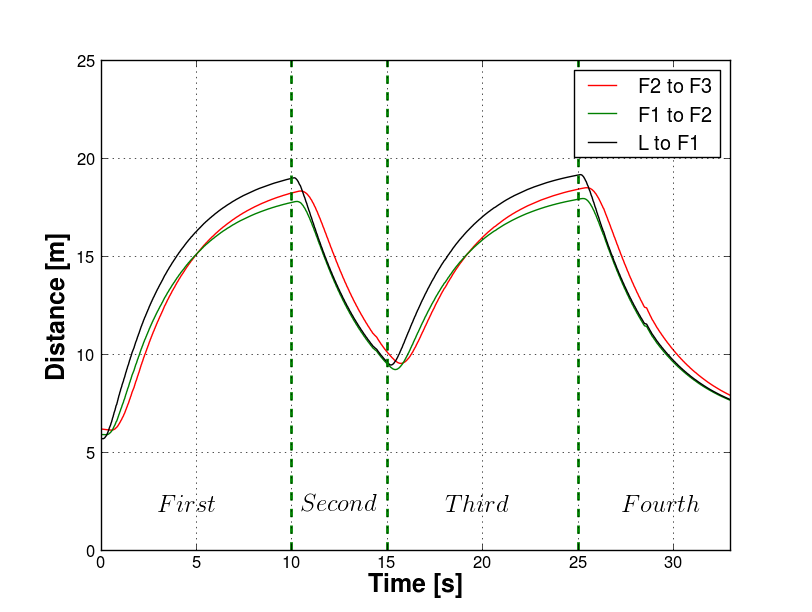}}}}
\caption{Velocity and distance between the vehicles with communication delay 0.01 s.}
\label{fig:control01}
\end{figure}

\begin{figure}[thpb]
\centering
\subfigure[Velocity in the $CACC$ system]{\framebox{{\includegraphics[width=7.3cm,height=4cm]{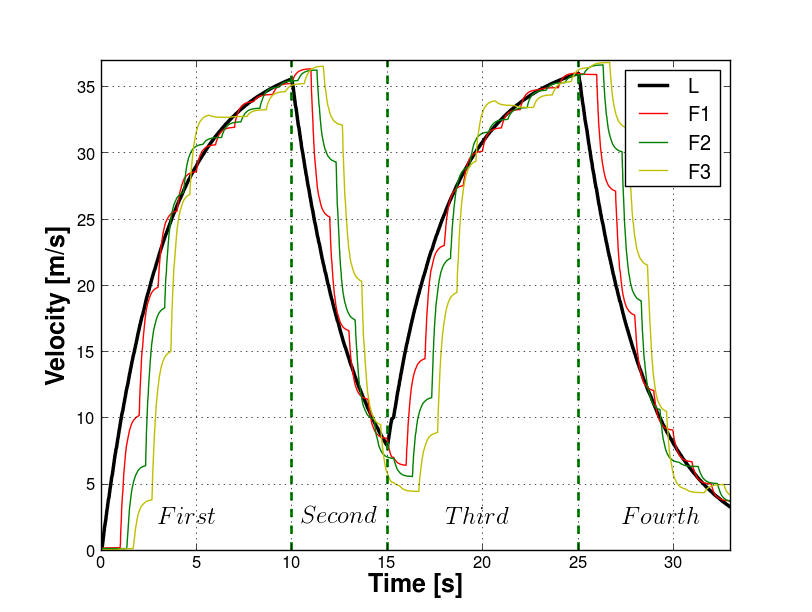}}}}
\subfigure[ Distance among vehicles]{\framebox{{\includegraphics[width=7.3cm,height=4cm]{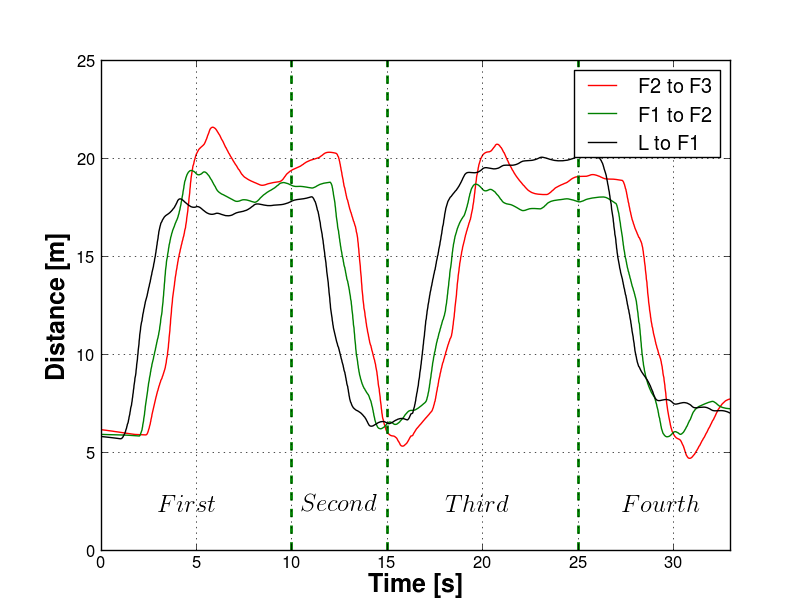}}}}
\caption{Velocity and distance between the vehicles with communication delay 0.333 s.}
\label{fig:control33}
\end{figure}

Figures \ref{fig:control01} and \ref{fig:control33} present the simulation results for communication delays of $0.01$ \textit{s} and $0.333$ \textit{s}. Figures \ref{fig:control01}a and \ref{fig:control33}a present the velocity of each vehicle over time for the two delay settings. In Figure \ref{fig:control01}a, it is possible to notice that when the leader is accelerating its velocity is higher than the velocity of the followers. During the leader deceleration we have the opposite behavior. It happens due to the small delay in the control response, which is caused by the inertia of the vehicle, communication delay, and other physical aspects of simulation. In Figure \ref{fig:control33}a, it is possible to observe how the behavior of the three followers is affected by the increment of the communication delay among vehicles. Consequently, an irregular behavior in each acceleration/deceleration pattern can be noticed.

Figure \ref{fig:control01}b and Figure \ref{fig:control33}b show the distance between the vehicles over time with communication delay of $0.01$ \textit{s} and $0.333$ \textit{s}, respectively. As stated in the $CACC$ control law, the distance varies according to the velocity of the vehicles. Nevertheless, it is possible to see that the distances between vehicles $L$$-$$F1$, $F1$$-$$F2$, and $F2$$-$$F3$ have the same profile. When the communication delay of $0.333$ \textit{s} was used, an increase in the difference of distance among vehicles of $CACC$ system was observed.

Studies consisted of experiments done in a real field-test, with similar characteristics to the experiments done with the proposed driving simulator platform, were used as a reference in order to contrast the results related to the effect of the time delay in the $CACC$ control. This decision is justified in Subsection \ref{ss:val_DS}.     

Among the different studies researched, the most valuable was the one by Ploeg et al. in \cite{Ploeg}, where the authors proposed a design and validation of a Cooperative Adaptive Cruise Control $CACC$ in a real field-test. In its experiment a fleet of six Toyota Prius III were used, since this model presents favorable technical characteristics for a real field-test. These vehicles were also equipped with other components in order to configure the $CACC$ system appropriately (see Section $IV$ in \cite{Ploeg}).     

Ploeg et al. considered the inclusion of time delay when developing the control implemented in the field-test (see section III-B in \cite{Ploeg}). According to the authors, the time delay may produce disturbances in the control of vehicles (i.e., string stability), which can also be observed in Figure \ref{fig:control33} of this study. For this reason, Ploeg et al. in \cite{Ploeg} considered a time delay of $150$ $ms$ in the communication system. The magnitude of such time delay was considered to be the best suited by the authors for their experiment. This value of time delay was also favorable to our results. Table \ref{t2} in Subsection \ref{ss:result} of this study, shows that a short time delay presents lower variations in the $CACC$ system. The obtained results in the $CACC$ validation, proposed by Ploeg et al. can be found in Section V-B of \cite{Ploeg}.  

\section{Influence evaluation of a Driving assistance system in a driver}\label{a2}

In order to evaluate the Driving Assistance System ($DAS$) influence on a driver, a test drive with twenty people was performed, considering two collision situations. In this evaluation, it was considered a driving simulator platform configuration that uses a \textit{DAS'} model with vehicular communication ($V2V$), sensors on the vehicles (e.g., $GPS$, $IMU$), a highway scenario, and the possibility to add an external hardware (i.e., \textit{Joystick G27}).  

\subsection{Driving Assistance System model}\label{ss:mDAS}

For this study, it was implemented the first case (i.e., \textit{following}) of the study proposed by Sebastian et al. in \cite{Sebastian}. In the test drive, the case \textit{following} occurs when two or more vehicles have the same orientation and direction on the same lane. Algorithm \ref{graph} illustrates the situation.

\begin{algorithm}
\caption{\textit{Case following}}\label{graph}
\begin{algorithmic}[1]
 	\For{\textbf{each} pair of vehicles ($A$, $B$)}\
	   \If{$|$$\theta_{A}$ $-$ $\theta_{B}$$|$ $<$ $\beta$}\
	   	 \State compute the distances $d_{p}$, $d_{ls}$ and $d_{a}$\
	     \State find the follower \textit{f} and the leader \textit{l}\
	     \State calculate safe distance \textit{$d_{sf}$}\
	     \If{$d_{a}$ $<$ $d_{sf}$}\
	       \State Collision warning 
	     \EndIf
       \EndIf 
	\EndFor
\end{algorithmic}
\end{algorithm}

\begin{figure}[thpb]
\centering
\framebox{\includegraphics[width=7cm,height=4.3cm]{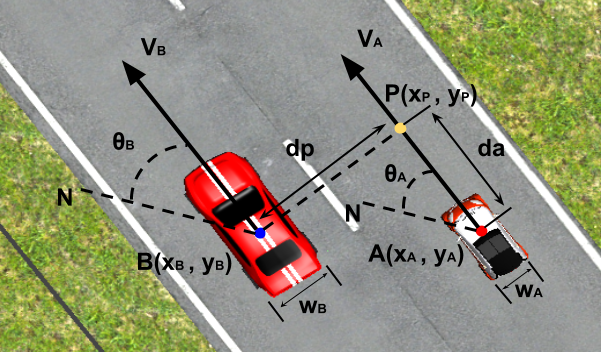}}
\caption{Vehicles with the same direction. Adapted from \cite{Sebastian}}
\label{fig:Cases}
\end{figure}

In the case \textit{following}, we have a vehicle \textit{A} following vehicle \textit{B} or vice versa. Figure \ref{fig:Cases} shows an example of case \textit{following}. The algorithm \textit{line 2} checks if the vehicles (i.e., \textit{A} and \textit{B}) are in the same orientation. This could be done by comparing the direction angles between the vehicles, in which $\theta_{A}$ $\approx$ $\theta_{B}$ (see Figure \ref{fig:Cases}). In the algorithm \ref{graph}, this comparison is determined as $|$$\theta_{A}$ $-$ $\theta_{B}$$|$ $<$ $\beta$, in which $\beta$ is a constant to be determined.%

Next in algorithm \textit{line 3}, the lateral distance $d_{p}$ and longitudinal distance $d_{a}$, between the $A$ and $B$ vehicles have to be computed. These distances are represented in Figure \ref{fig:Cases}, and may be calculated using equations \ref{eq:5} and \ref{eq:6}. However, to get equation \ref{eq:5} is necessary to consider the value of the next variables:%
\begin{equation}
{ x }_{ A' } = \sin { { \theta  }_{ A } } +{ x }_{ A }\quad ,\quad { y }_{ A' } = \cos { { \theta  }_{ A } } +{ y }_{ A }
\label{eq:1}
\end{equation}%
\begin{equation}
u=\frac { \left( { x }_{ B }-{ x }_{ A } \right) \left( { x }_{ A' }-x_{ A } \right) +\left( y_{ B }-{ y }_{ A } \right) \left( { y }_{ A' }-{ y }_{ A } \right)  }{ \sqrt { { \left( { x }_{ A' }-{ x }_{ A } \right)  }^{ 2 }+{ \left( { y }_{ A' }-{ y }_{ A } \right)  }^{ 2 } }  }
\label{eq:2}
\end{equation}%
\begin{equation}
{ x }_{ P }\quad =\quad { x }_{A}+u\left( { x }_{ A' }-x_{ A } \right)
\label{eq:3}
\end{equation}%
\begin{equation}
{ y }_{ P }\quad =\quad y_{ A }+u\left( { y }_{ A' }-{ y }_{ A } \right) 
\label{eq:4}
\end{equation}%
\begin{equation}
{ d }_{ p }\quad =\quad \sqrt { { \left( { x }_{ P }-{ x }_{ B } \right)  }^{ 2 }+{ \left( { y }_{ P }-{ y }_{ B } \right)  }^{ 2 } } 
\label{eq:5}
\end{equation}%
The longitudinal distance $d_{a}$ is also calculated based on the Euclidean Distance between the $A$ vehicle's position and the found coordinate \textit{P} (see Figure \ref{fig:Cases}) using equations \ref{eq:1},\ref{eq:2},\ref{eq:3} and \ref{eq:4}:%
\begin{equation}
{ d }_{ a }\quad =\quad \sqrt { { \left( { x }_{ P }-{ x }_{ A } \right)  }^{ 2 }+{ \left( { y }_{ P }-{ y }_{ A } \right)  }^{ 2 } } 
\label{eq:6}
\end{equation}%
On the other hand, it is also necessary to compute the lateral safety distance ${d_{ls}}$ with the equation \ref{eq:7}. The distance ${d_{ls}}$ allows to decide if the two vehicles are in the same lane, considering the width of $A$ vehicle ($\omega_{A}$), the width of $B$ vehicle ($\omega_{B}$), and a minimal lateral safety distance ${{d}_{mls}}$ that have to be defined.%
\begin{equation}
{d}_{ls}\quad =\quad \frac { w_{A} }{ 2 } + \frac { w_{B} }{ 2 } +d_{mls}
\label{eq:7}
\end{equation}%
In order to find out if a vehicle is a \textit{leader} or a \textit{follower} (i.e., algorithm's \textit{line 4}), the original algorithm had some changes. In this new approach, it is defined a vector $\overrightarrow{AP}$ between the points $A$ and $P$ (see Figure \ref{fig:Cases}), after it is found the angle $\phi$ between the vectors $\overrightarrow{AP}$ and $\overrightarrow{V_{A}}$ (i.e., $\overrightarrow{V_{A}}$ is the $A$ velocity vector), using equation \ref{eq:8}:%
\begin{equation}
\phi = \arccos { \frac { \overrightarrow{AP}\cdot \overrightarrow{V_{A}} }{ \left| \overrightarrow{AP} \right| \left| \overrightarrow{V_{A}} \right|  }  } 
\label{eq:8}
\end{equation}
 
If $\phi < \pi/2$ the $A$ vehicle is a \textit{follower} otherwise it is a \textit{leader}. Finally, it is calculated the safety distance ${ d }_{ sf }$ using Equation \ref{eq:9} (i.e., algorithm's \textit{line 5}).%
\begin{equation}
{ d }_{ sf }={ d }_{ min }+{ v }_{ f }{ t }_{ r }+\frac { 1 }{ 2 } \left( \frac { { v }_{ f }^{ 2 } }{ { a }_{ f } } -\frac { { v }_{ l }^{ 2 } }{ { a }_{ l } }  \right) 
\label{eq:9}
\end{equation}%
In which $d_{min}$ is the minimal distance between $A$ and $B$ vehicles to avoid a collision; $v_{f}$ is the following vehicle speed; $t_{r}$ is a parameter of driver's reaction time; $a_{f}$ is the following vehicle maximum deceleration; $v_{l}$ is the leading vehicle speed and $a_{l}$ is the leading vehicle maximum deceleration. The value of the distances $d_{a}$ and $d_{sf}$ are necessary to determine a \textit{Collision Warning} condition (i.e., algorithm \textit{line 6}).

\subsection{Driving Simulator Platform Configuration}\label{cap:DSPC2}

\begin{figure}[thpb]
\centering
\framebox{\includegraphics[width=8cm,height=5cm]{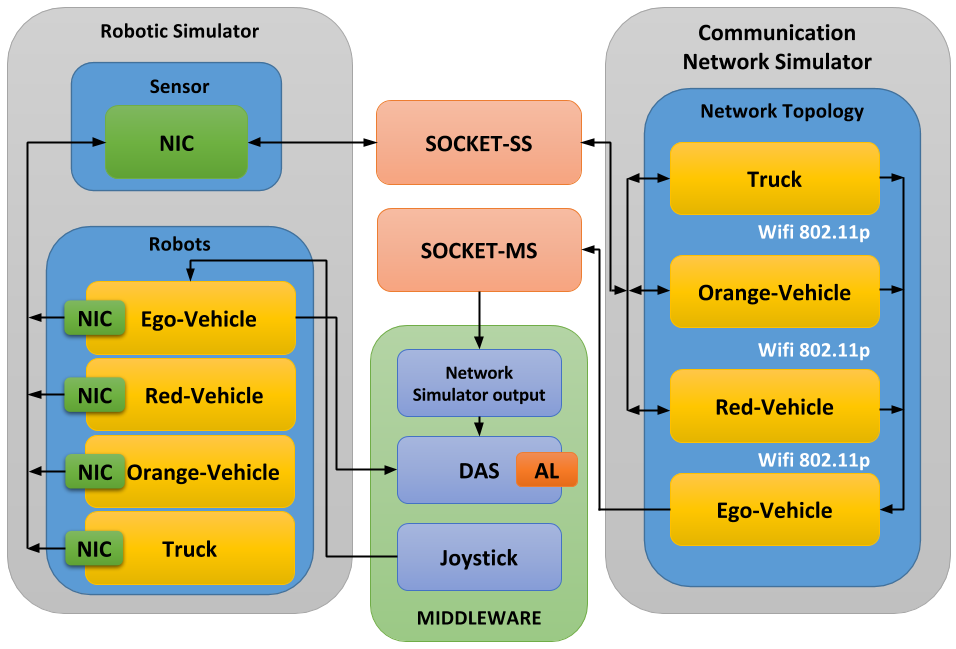}}
\caption{Driving simulator platform configuration used to evaluate the \textit{DAS'} influence in a driver.}
\label{fig:arq2}
\end{figure}%
Figure \ref{fig:arq2} presents the configuration of the driving simulator platform's architecture. Its new configuration was the result of carefully planned experiments to validate the selected \textit{DAS} model influence on a driver. These experiments (see Subsection \ref{V_exp}) required an interaction among vehicles, a highway scenario, a new $V2V$ communication network topology and the use of a joystick that would allow the interaction of a user with the driving simulator platform.    

Inside of the rectangle \textit{Robotic Simulator}, which represents the $MORSE$ simulator in Figure \ref{fig:arq2}, it is possible to observe the representation of four different vehicles. For the $3D$ model of the vehicles in $MORSE$, it was considered the use of sensors (i.e., GPS, velocity, pose and \textit{NIC}) as it is presented in Subsection \ref{cap:DSPC1}.    

For this new driving simulation platform configuration, in the message of the sensor Network Interface Communication \textit{NIC}, it was considered a new value. This new value is the distance that exists between the \textit{Ego-vehicle} and each vehicle present inside the highway scenario of the simulation. The distance is calculated in the sensor \textit{NIC} using the vehicles’ location data. This information is important because this value determines the communication range of the vehicles in the new communications model implemented in $NS-3$.  

The rectangle \textit{Communication Network simulator}, that represents the $NS-3$ simulator in Figure \ref{fig:arq2}, obtains the sent messages by  the sensor \textit{NIC}, again through \textit{SOCKET-SS} (see Section \ref{s:III}). The communications model implemented in $NS-3$ represents, through nodes, each one of the vehicles used in $MORSE$. The structure of the $V2V$ communication model allows the node \textit{Ego-vehicle} to receive messages from the other vehicles’ nodes present in the simulation scenario. If the distance value among the vehicles (i.e., obtained in the sent message by \textit{NIC}) is greater than a range of $300$ \textit{m}, the communications model determines that the node \textit{Ego-vehicle} cannot receive messages. When other vehicles enter the communication range of the \textit{Ego-vehicle}, the communication begins. The $V2V$ communications model do not allow the node \textit{Ego-vehicle} to sent messages to the others vehicles’ nodes, as it can be observed in the rectangle \textit{Communication Network simulator} in Figure \ref{fig:arq2}.     

When the \textit{Ego-vehicle} node receives a message, from the node of another vehicle, the \textit{Ego-vehicle's} node sends this information for the rectangle \textit{MIDDLEWARE}, through \textit{SOCKET-MS}. In Figure \ref{fig:arq2}, the rectangle \textit{MIDDLEWARE}, which represents $ROS$ is composed by the nodes: \textit{i)} Network Simulator output; \textit{ii)} Driving Assistance System (\textit{DAS}); and \textit{iii)} Joystick (i.e., A \textit{ROS} system is comprised of a number of independent nodes, each of which communicates with the other nodes using a publish/subscribe messaging model. The \textit{ROS'} nodes are different of nodes used in $NS-3$). 

The node \textit{Network Simulator output} receives a message obtained from \textit{SOCKET-MS} and makes it available for the $ROS'$ system. The \textit{DAS} node is the piece of software in the $ROS'$ system where it is implemented the model presented in the Subsection \ref{ss:mDAS}. For the model execution, the \textit{DAS} node needs the data obtained from the \textit{Network Simulator output} node and sensor data from \textit{Ego-vehicle}, which are obtained from $MORSE$ (see Figure \ref{fig:arq2}). $MORSE$ allows a direct communication with $ROS$. Finally, the \textit{joystick} node allows the interaction between the user’s and the driving simulator platform. In this node are implemented the configuration and the execution of the joystick functions, needed in the experiments. As it can be observed in Figure \ref{fig:arq2}, this node also communicates directly with the \textit{Ego-vehicle} in $MORSE$. 


\subsection{Experiments}\label{V_exp}

With the objective to validate how the implemented $DAS'$ model influences in a driver, two of thirty-seven potential collision situations were chosen (i.e., a lead vehicle stops and a vehicle changing lanes in the same direction). These collision situations are proposed by the National Highway Traffic Safety Administration (\textit{NHTSA}) in \cite{NHTSA}.  Those collision situations are the result of several crash analysis, based on the accidents data collected in the United States in \textit{2009}, which could have been avoided using a system with $V2V$ communications technology. From this result, the risk typology in this section was developed.  In Subsection \textit{III-A} of \cite{NHTSA} it can be observed the other thirty-five collision situations.  

For the evaluation of the $DAS'$ model influence on a driver,postgraduate students from the \textit{science institute of mathematics and computers} (\textit{ICMC-USP}) were invited to participate in the experiment. It was not disclosed to the participants the main purpose of the experiment, and each participant was required to have a valid driver’s license and driving experience of a manual gear shift. Only twenty postgraduate students voluntarily accepted the invitation to participate. From the $20$ participants $7$ were female and $13$ were male, and the participants' age averaged $27.2$ years (i.e., sd = $3.65$ years). They were divided into two groups: \textit{i)} without $DAS$ (i.e., $4$ women, and $7$ men); and \textit{ii)} with $DAS$ (i.e., $3$ women, $6$ men), in order to verify the $DAS’$ model influence in the two collision situations the drivers from both groups were exposed to.    

For the implementation of the two collision situations, it was developed a virtual highway based on a specific segment of the highway \textit{Washington Luiz}, in the S\~ao Paulo state (i.e., from the bridge at the main access way to S\~ao Carlos Street until the u-turn at exit $240$). In this virtual highway, the $GPS$ coordinates of the real highway segment were used in order to obtain the real shape of the highway structure. Moreover, $Blender’s$ tools were also used (see Section \ref{s:III}) to add some characteristics in the virtual highway, such as: \textit{i)} asphalt under ideal conditions, \textit{ii)} the grass around the highway, \textit{iii)} the use of objects and infrastructure present in the real highway. For the virtual highway scenario, it was considered a sunny environment without climatological changes and one-way lanes.

Four vehicles were used in the virtual highway scenario. Only one of these vehicles was driven by the participants (i.e., \textit{Ego-vehicle}) using the \textit{joystick G27}, while the other three vehicles were moving autonomously. For the experiment, each participant received instructions on how to operate the \textit{joystick G27}, which consisted of a steering wheel, accelerator, brake and clutch for the gear shift (See Figure \ref{fig:platform}). Before the experiment began and data were collected, the participants had time to adapt to the driving simulator platform.     

The three vehicles that were moving autonomously were controlled by an implemented algorithm, which defined their paths within the virtual scenario. Two of these vehicles were used on the collision situations (i.e., a conventional red vehicle and a truck), and the third vehicle was used to distract the drivers (i.e., a conventional orange vehicle). The sequence of events of the simulation consisted of the first collision situation, the encounter of the drivers with the orange vehicle and the second collision situation. As follows, is a detailed description of the characteristics considered for each collision situation:

\subsubsection{The lead vehicle stopped} to configure the danger in this situation, it was developed a routine in the algorithm for the red vehicle moving autonomously. In this routine, the red vehicle detects the presence of the \textit{Ego-vehicle} in a range of $30$ \textit{m} from its rear. It then activates the brakes in the vehicle controller, used by $Blender$ in the red vehicle model (see Section \ref{ss:dynamic}), instantaneously decelerating the red vehicle until it is completely stopped.

When a participant starts the experiment, the \textit{Ego-vehicle} is parked on the virtual highway. After the participant drives for a while in the virtual scenario path, the user encounters the other three vehicles. Two of these vehicles, the red and orange, are moving in each lane of the highway (i.e., highway with just two lanes), at a velocity of $70$ \textit{km/h}, which prevents an overtaking maneuver between them. This is the moment the red vehicle detects the \textit{Ego-vehicle} and the first collision situation initiates.  

In this collision situation, the participant from the group using the \textit{DAS’} model receives a warning sound, which serves to alert the participant of the collision risk that the red vehicle ahead could represent. The sound signal is enabled when the condition in line six of Algorithm \ref{graph} is accomplished. In Figure \ref{fig:arq2} it is possible to observe the relationship between the node $DAS$ in $ROS$ with the sound signal (i.e., \textit{AL}).    
     
\subsubsection{A vehicle changing lanes in the same direction} In this second collision situation, it was used a $3D$ model of a truck as one of the autonomous vehicle in the simulation. Here, the algorithm implemented for the truck first percepts (i.e., in a range of $15$ \textit{m}) the presence of the \textit{Ego-vehicle} approaching behind of it in a different lane. Then, the routine in the algorithm activates the steering control of the vehicle controller, used by $Blender$ in the truck model, causing the truck to move towards the \textit{Ego-Vehicle's} lane. The sudden change of lane by the truck, moving at $70$ \textit{km/h} on the highway, initiates the second collision situation of the experiment. Like in the first situation, a warning sound alerts the participant from the group using \textit{DAS} of the truck’s presence.      

The orange vehicle’s sole purpose is to distract the attention of the driver between the two collision situations, but without causing any further risks to the driver, who overtakes the orange vehicle without difficulties. A video of these experiment is available on \url{https://www.youtube.com/watch?v=H4lXVlq_8lI}. 

\subsection{Results}

To obtain an experience feedback, when the platform users completed the experiment, each participant completed a survey containing seven questions. With the survey results was possible to observe a worrying situation from two of the questions. In \textit{Did you break the speed limit?} 65\% of the users answered affirmatively. In \textit{Did you make some dangerous maneuvers during the trial route?} 70\% of the users also answered affirmatively. This result confirms that drivers between 21 and 30 years of age have more than 50\% chance to cause an accident, as also seen in Stiller et al. \cite{stiller}, where it analyzed the driving assistance systems impact in the last thirty years.%
\begin{figure}[!htb]
\centering
\subfigure[Did you crash with some vehicle?]{\framebox{{\includegraphics[width=7cm,height=4.5cm]{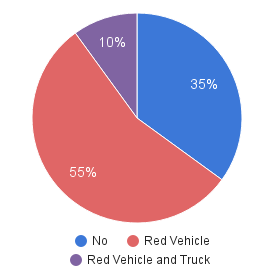}}}}
\subfigure[Did you think that the Driver Assistance System helped in the test?]{\framebox{{\includegraphics[width=7cm,height=4.5cm]{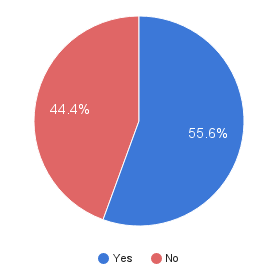}}}}
\caption{Survey questions}\label{fig:sur}
\end{figure}%

Figure \ref{fig:sur} presents the results of two other questions from the survey. Figure \ref{fig:sur}a shows that the most dangerous situation for the driving simulator platform users was the first one (i.e., the lead vehicle stopped). Figure \ref{fig:sur}b, in relation to the question \textit{Did you think that the Driver Assistance System helped in the test?}, in the group with $DAS$, only $56$\% of the users considered the $DAS'$ model useful in the test.

To evaluate the $DAS’$ influence in a driver, the driving simulator platform made a \textit{test log}, with the objective to obtain some data of each user during the experiment, described in Subsection \ref{V_exp}. The data obtained in \textit{log} are \textit{Ego-vehicle's} velocity, the \textit{Ego-vehicle's} brake application rate, and the initial time of the collision situation.

\begin{figure}[tbp]
\centering
\subfigure[Velocity effect in the two collision situations]{\framebox{{\includegraphics[width=8.3cm,height=5cm]{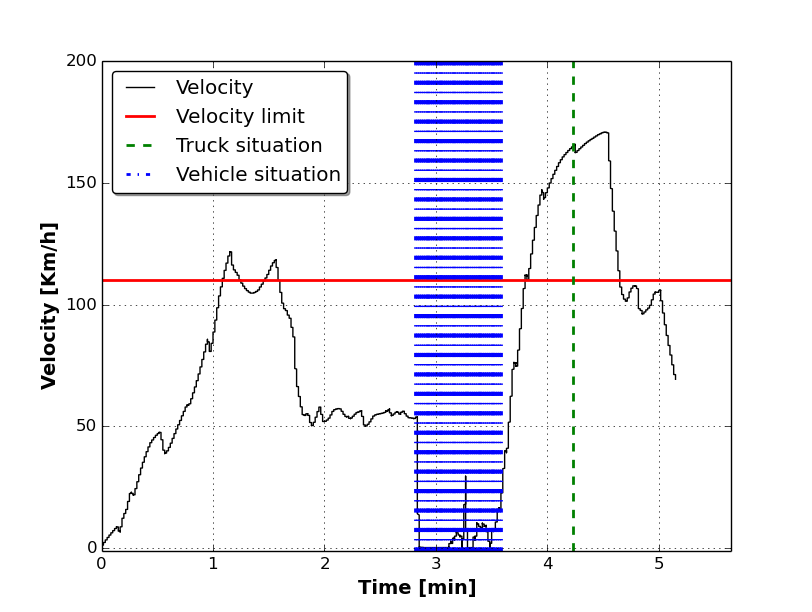}}}}
\subfigure[Brake effect in the two collision situations]{\framebox{{\includegraphics[width=8.3cm,height=5cm]{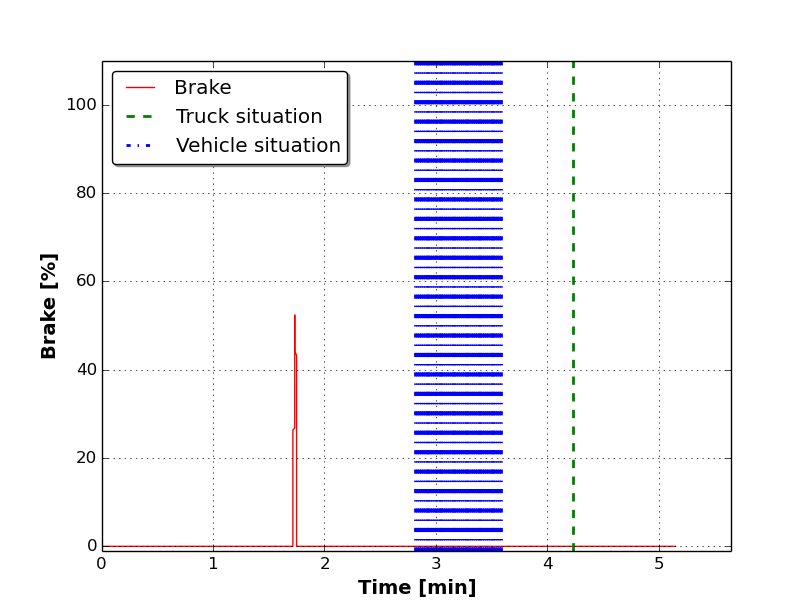}}}}
\caption{A participant's log without DAS' model.}\label{fig:wDAS}
\end{figure}

\begin{figure}[tbp]
\centering
\subfigure[Velocity effect in the two collision situations]{\framebox{{\includegraphics[width=8.3cm,height=5cm]{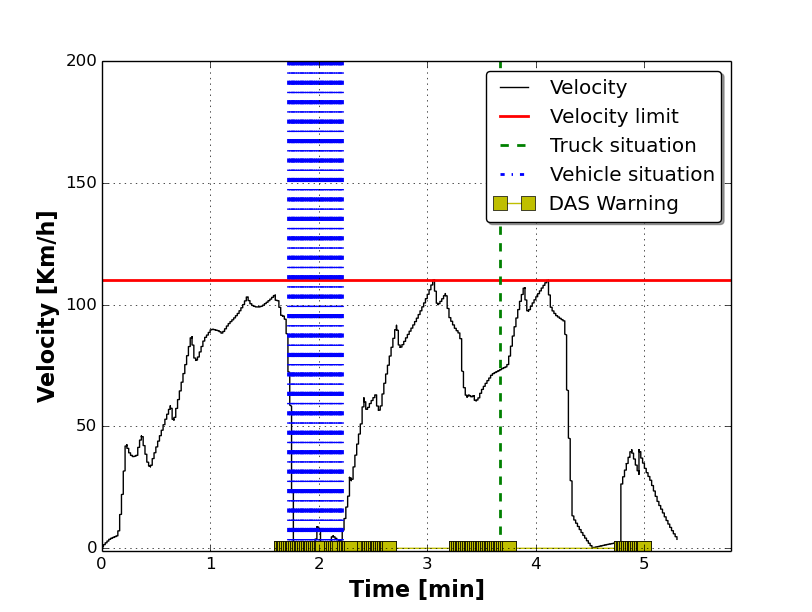}}}}
\subfigure[Brake effect in the two collision situations]{\framebox{{\includegraphics[width=8.3cm,height=5cm]{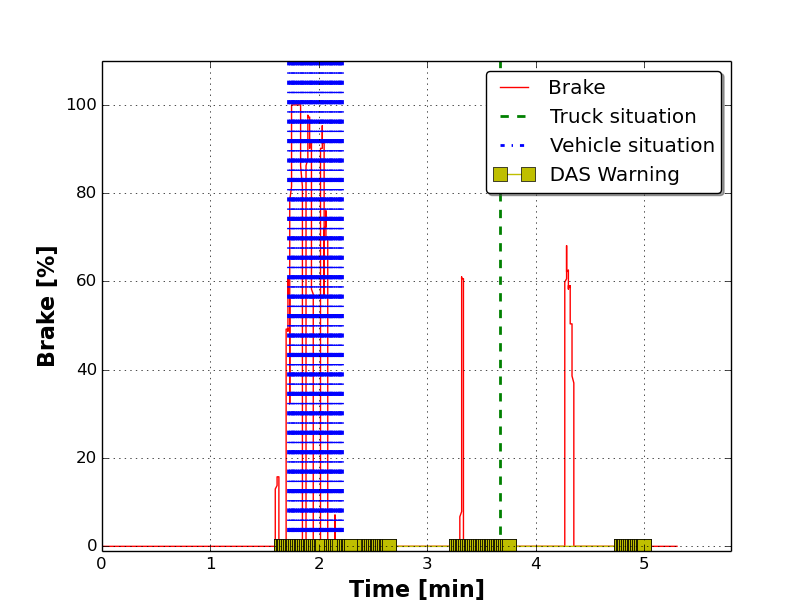}}}}
\caption{A participant's log with DAS' model.}\label{fig:das}
\end{figure}

The initial time's value of the collision situation helps to find the magnitude of the \textit{Ego-vehicle's} velocity and the \textit{Ego-vehicle's} brake application rate in the exact moment, when beginning the collision situation inside of each user’s \textit{log} (see vehicle situation and the truck situation in Figures \ref{fig:wDAS} and \ref{fig:das}). The \textit{Ego-vehicle's} velocity, and the \textit{Ego-vehicle's} brake application rate are the only data considered to analyze the $DAS’$ influence in drivers.

In order to analyze the results, were considered the mean, median, standard deviation, and the maximum and minimum values from the number of samples (i.e., participants) in the experiment by each group (see the second paragraph in Subsection \ref{V_exp}). In Tables \ref{t:twtd} and \ref{t:twd} can be observed this statistical parameters. In addition to these parameters, it was used a \textit{bootstrap method} to determine a $95$\% confidence intervals (i.e., $Bmin$ and $Bmax$ in Tables \ref{t:twtd} and \ref{t:twd}), with the objective to compare the data set of \textit{Ego-vehicle's} velocity and the \textit{Ego-vehicle's} brake application rate for each group. The \textit{bootstrap method} is used for the results analysis, with small sample drawn from a distribution that differs from a normal distribution. In this approach, no assumptions were made about the original population \cite{Triola}.

\begin{table}[tbp]
\centering
\caption{Results group without \textit{DAS}}
\label{t:twtd}
\begin{tabular}{|c|c|c|c|c|}
\hline
\textbf{Parameter} & \textbf{VCS1 (Km/h)} & \textbf{BCS1(\%)} & \textbf{VCS2 (Km/h)} & \textbf{BCS2(\%)} \\ \hline
\textbf{Mean}      & 68.98                & 11.32             & 95.07                & 0.35              \\ \hline
\textbf{Median}    & 66.43                & 0.0               & 92.71                & 0.0               \\ \hline
\textbf{Std}       & 17.02                & 24.49             & 28.86                & 1.18              \\ \hline
\textbf{Min}       & 42.95                & 0.0               & 54.59                & 0.0               \\ \hline
\textbf{Max}       & 101.72               & 80.0              & 165.33               & 3.94              \\ \hline
\textbf{BD min}    & 60.28                & 1.75              & 81.98                & 0.0               \\ \hline
\textbf{BD max}    & 79.61                & 32.74             & 116.41               & 1.07              \\ \hline
\end{tabular}
\end{table}

By comparing the statistical parameters’ values shown in Tables \ref{t:twtd} and \ref{t:twd}, for the \textit{Ego-vehicle's} velocity data (i.e., $VCS1$, \textit{Ego-vehicle's} velocity in the first collision situation), it is possible to observe that the mean, the median, and the bootstrap’s confidence intervals  (i.e., $Bmax$ and $Bmin$) in the two groups do not have significant differences.

Moreover, for the \textit{Ego-vehicle's} brake application rate data (i.e., $BCS1$, \textit{Ego-vehicle's} brake application in the first collision situation), the values of the median, in Tables \ref{t:twtd} and \ref{t:twd}, show that the brake practically was not activated in both groups since the most probable value in the median was \textit{zero}. This fact shows that, during the first collision situation, a considerable majority of the experiment participants, in both groups, did not react in the beginning of the collision situation.  

\begin{table}[tbp]
\centering
\caption{Results group with \textit{DAS}}
\label{t:twd}
\begin{tabular}{|c|c|c|c|c|}
\hline
\textbf{Parameter} & \textbf{VCS1 (Km/h)} & \textbf{BCS1(\%)} & \textbf{VCS2 (Km/h)} & \textbf{BCS2(\%)} \\ \hline
\textbf{Mean}      & 66.03                & 7.57              & 73.32                & 0.0               \\ \hline
\textbf{Median}    & 62.11                & 0.0               & 77.91                & 0.0               \\ \hline
\textbf{Std}       & 10.01                & 19.96             & 27.74                & 0.0               \\ \hline
\textbf{Min}       & 57.44                & 0.0               & 34.57                & 0.0               \\ \hline
\textbf{Max}       & 87.91                & 60.62             & 113.8                & 0.0               \\ \hline
\textbf{BD min}    & 61.30                & 0.52              & 55.98                & None              \\ \hline
\textbf{BD max}    & 74.59                & 28.61             & 89.99                & None              \\ \hline
\end{tabular}
\end{table}%

Another aspect observed in Tables \ref{t:twtd} and \ref{t:twd} for $BCS1$, was that the bootstrap’s confidence intervals (i.e., $Bmax$ and $Bmin$) do not present a considerable difference. Therefore, these values can determine that the use of the $DAS’$ model, during the first collision situation did not have an impact on the driver. In Figure \ref{fig:sur}a is possible to observe this fact, given that it was the red vehicle (i.e., used in the experiment to generate the first collision situation) that, according to the survey completed by the experiment participants, caused the majority of the crashes.     

On the other hand, in the second collision situation, the differences between the statistical parameters' values, in Tables \ref{t:twtd} and \ref{t:twd} are considerable, for the \textit{Ego-vehicle's} velocity data (i.e., $VCS2$, \textit{Ego-vehicle's} velocity in the second collision situation). Only the standard deviation between the groups has a slight difference. The most considerable difference between the two groups can be found between the bootstrap’s confidence intervals for $VCS2$. 

Therefore, with a $95$\% of confidence it is possible to evidence that the participants’ group using the $DAS’$ model was moving at approximately $20$ \textit{km/h} slower than the other participants’ group which did not use the $DAS’$ model, in the beginning of the second collision situation. Figures \ref{fig:wDAS} and \ref{fig:das} represent this fact, considering the relation between the labels, velocity (i.e., \textit{Ego-vehicle} velocity) and truck situation. 

Comparing the statistical parameters obtained for the \textit{Ego-vehicle's} brake application rate data (i.e., $BCS2$, \textit{Ego-vehicle's} Brake application in the second collision situation), it was possible to verify that all the participants of the group that used the $DAS’$ model, did not used the \textit{Ego-vehicle’s} brake during the beginning of the second dangerous situation. 

Consequently, after analyzing the experiment’s results, it is possible to determine that the $DAS’$ model, in the second collision situation, could have an influence on the driver. Figure  \ref{fig:das}b is an example of a participant’s log from the group that used the $DAS’$ model. In this figure is possible to observe that \textit{Ego-vehicle's} brake was activated seconds after that the $DAS’$ model adverted the driver.    

Finally, considering that the two collision situations were consecutive, it could be determined that the experiment’s participants, who used the $DAS’$ model, understood the model’s effect only after the first collision situation. Therefore, although the $DAS$ model works, it could be thought that the $DAS$ model does not have the level of influence necessary for a driver to avoid an accident.


\section{Conclusion}

For the driving systems development, the use of a driving simulator platform can facilitate the work for the design time as well as the economic, temporal and logistical aspects, among others. However, with the obtained results, the use of the driving simulator platform cannot be considered as definitive. The simulation tools should only be used in a starting level. 

A simulation tool could have best results if the models used in the simulation platforms (e.g., sensors, vehicles, communications and so on) have a feedback from obtained results in the field-test. Nevertheless, this feedback requires flexible driving simulation platforms (i.e., mainly in research) that allow the researcher to modify the characteristics in the setup of the driving simulation platform.

Taking into account these question, in this study, it is presented an adaptable driving simulator platform that allows the platform users to adjust its characteristics (e.g., hardware and software), for an experimental implementation, regarding technical and economic requirements. The two applications presented in Sections \ref{a1} and \ref{a2} allow us to prThis decision is justified by Engen inove the flexibility of the driving simulator platform.  

The control system development, for the autonomous vehicle navigation, is an example of the simulation platform validation. With the use of the proposed driving simulator platform was possible to develop and evaluate the control system, in starting level, and to optimize the available resources.  

The use of the driving simulator platform has allowed to analyze the influence of the $DAS$ on a driver. Considering the results, it is clear that the use of an informative approach (i.e., a warning sound) to alert the driver about the danger is not enough. Therefore, although the $DAS$ model works, This model does not guarantee that the driver can avoid an accident. 

For future work, the next version of the Vehicular Dynamic Model (\textit{VDM}) in the driving simulator platform has to consider some aspects, such as: \textit{i)} to edit and easily replace, the different components of the \textit{VDM} , \textit{ii)} to use the main parameters of real vehicles, in the vehicles' model design, and finally, \textit{iii)} the model must be developed to be stiff and efficient from a numerical point of view. It is also important to mention that the simulation platform will be available to other research groups, with the objective to enable a comparison between this proposal and other approaches. 

\section*{Acknowledgment}

The authors acknowledge the support granted by CNPq and CAPES.

\ifCLASSOPTIONcaptionsoff
  \newpage
\fi


%
%

%
\begin{IEEEbiography}[{\includegraphics[width=2.8cm,height=3.3cm]{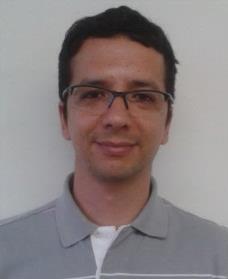}}]{Andr\'es E. G\'omez H.} received the Engineering degree in Electronic from the University of Pamplona ($UP$), Pamplona, Colombia, in 2003, and the Master degree in Engineering and computer systems from the Industrial University of Santander ($UIS$), Bucaramanga, Colombia, in 2012. Currently he is Phd student of the sciences institute of mathematics and computers - University of S\~ao Paulo ($USP$). His current research interests are Driving Assistance Systems, Vehicular Communications, Intelligent Transportation Systems, Driving Simulators. 
\end{IEEEbiography}%
\begin{IEEEbiography}[{\includegraphics[width=2.8cm,height=3.3cm]{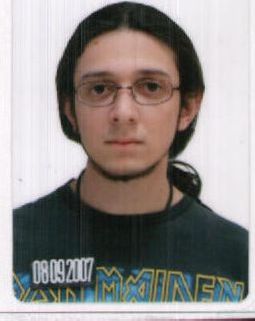}}]{Tiago C. dos Santos}received the Computer Science degree and master degree from the University of S\~ao Paulo ($USP$). Currently he is Phd student at Mobile Robotics Laboratory located in S\~ao Carlos - SP, Sciences Institute of Mathematics and Computers - ($USP$).
\end{IEEEbiography}%
\begin{IEEEbiography}[{\includegraphics[width=2.8cm,height=3.3cm]{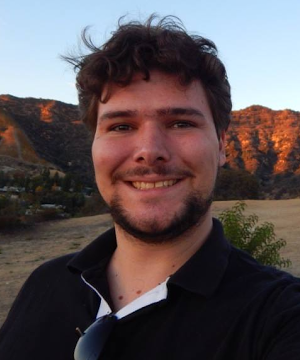}}]{Carlos M. Massera} is currently working towards his Ph.D. degree at the institute of Mathematics and Computer Science at the University of São Paulo. He received his B.Sc. in Computer Engineering from São Carlos School of Engineer in 2012. He has been working on autonomous and cooperative vehicle control and estimation systems and his current research interests are constrained optimal control, robust optimal control, robust convex optimization and its applications to autonomous and cooperative vehicles.
\end{IEEEbiography}%
\begin{IEEEbiography}[{\includegraphics[width=2.8cm,height=3.3cm]{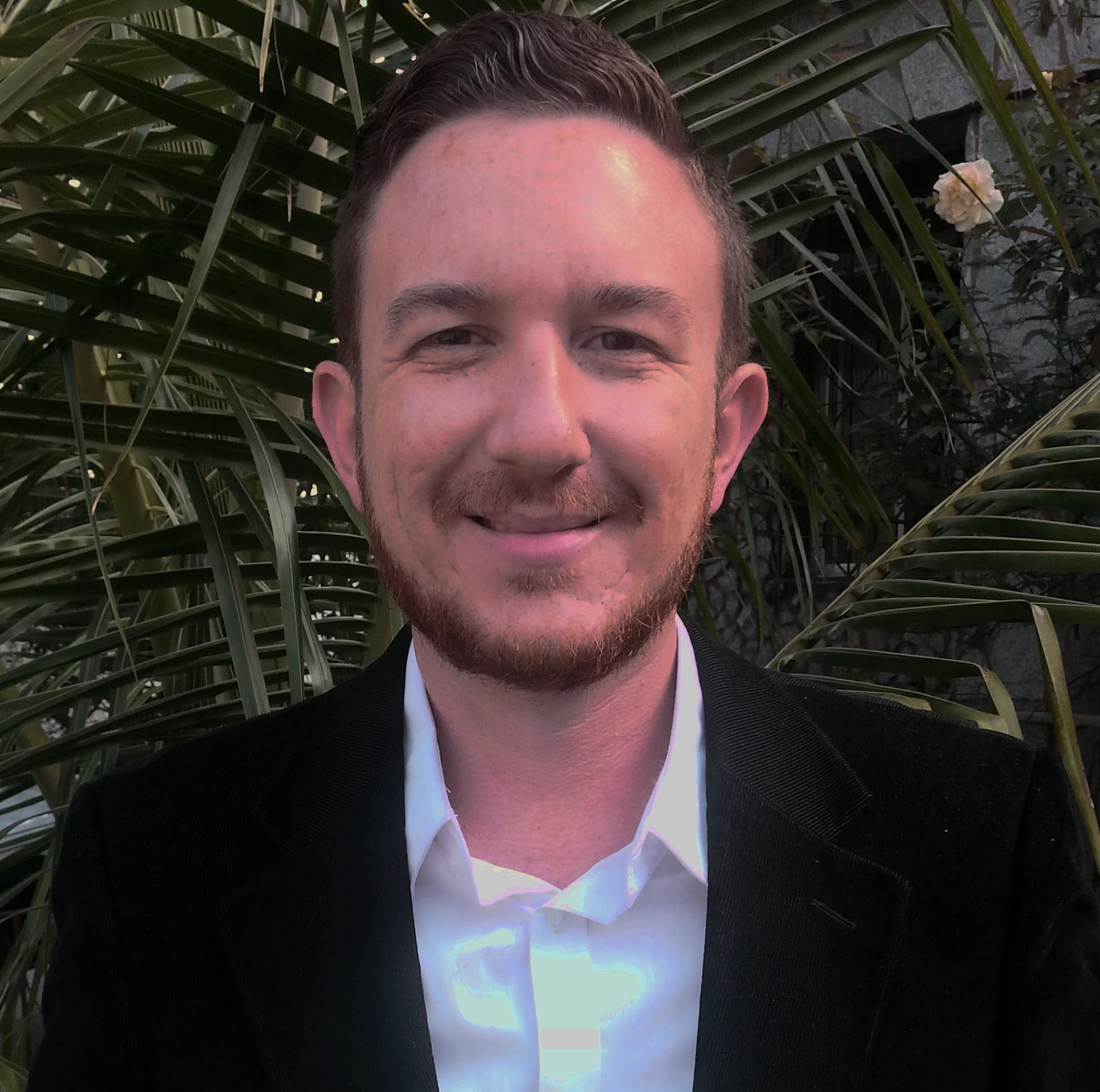}}]{Arthur de M. Neto} received the B.S. degree in computer science - data processing - (1998) and M.S. degree in mechanical engineering (2007) from State University of Campinas (UNICAMP), Campinas, Brazil, and a Ph.D. degree in information and systems technologies (2011) from the University of Technology of Compiegne (UTC), Compiegne, France, and in mechanical engineering from the UNICAMP, Brazil. Since 2014, he is an Associate Professor in the Engineering Department at the Federal University of Lavras (UFLA) and member (Head) of the Terrestrial Mobility Laboratory (LMT). His research interests are in the field of Robotic and Mechatronic Systems, more precisely in the area of Intelligent and Robotic Vehicles, the Development of Autonomous Navigation and Driver Assistance Systems, and Exteroceptive Perception of the Environment.
\end{IEEEbiography}%
\begin{IEEEbiography}[{\includegraphics[width=2.8cm,height=3.3cm]{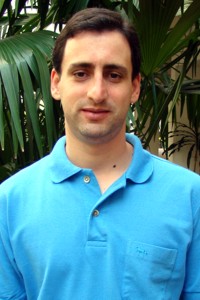}}]{Denis F. Wolf} is an Associate Professor in the Department of Computer Systems at the University of Sao Paulo (ICMC-USP). He obtained his PhD degree in Computer Science at the University of Southern California USC in 2006. Currently he is Director of the Mobile Robotics Laboratory at ICMC/USP and board member of the USP Center for Robotics. His current research interests are Mobile Robotics, Intelligent Transportation Systems, Machine Learning and Computer Vision.
\end{IEEEbiography}





\end{document}